\documentclass[acmsmall, authorversion]{acmart}

\usepackage{amsmath}
\usepackage{booktabs}
\usepackage{array}
\usepackage{multirow}
\usepackage{url}
\usepackage{breakurl}
\usepackage{balance}
\usepackage{makecell}
\usepackage{colortbl}
\usepackage{color}
\usepackage{tikz}
\usetikzlibrary{shapes, backgrounds}

\definecolor{highlight}{rgb}{0.94,0.94,0.94}

\AtBeginDocument{%
  }

\setcopyright{cc}
\setcctype{by}
\acmJournal{PACMHCI}
\acmYear{2025} \acmVolume{9} \acmNumber{7} \acmArticle{CSCW301} \acmMonth{11} 
\acmDOI{10.1145/3757482}

\begin{document}

\title[Impact Assessment Card]{Impact Assessment Card: Communicating Risks and Benefits of AI Uses}

\author{Edyta Bogucka}
\orcid{0000-0002-8774-2386}
\affiliation{
  \institution{Nokia Bell Labs}
  \city{Cambridge}
  \country{United Kingdom}}
\affiliation{
  \institution{University of Cambridge}
  \city{Cambridge}
  \country{United Kingdom}}
\email{edyta.bogucka@nokia-bell-labs.com}

\author{Marios Constantinides}
\orcid{0000-0003-1454-0641}
\affiliation{
  \institution{CYENS Centre of Excellence}
  \city{Nicosia}
  \country{Cyprus}}
\affiliation{
  \institution{Nokia Bell Labs}
  \city{Cambridge}
  \country{United Kingdom}}
\email{marios.constantinides@cyens.org.cy}

\author{Sanja \v{S}\'{c}epanovi\'{c}}
\orcid{0000-0000-0000-0000}
\affiliation{
  \institution{Nokia Bell Labs}
  \city{Cambridge}
  \country{United Kingdom}}
\affiliation{
  \institution{University of Oxford}
  \city{Oxford}
  \country{United Kingdom}}
\email{sanja.scepanovic@nokia-bell-labs.com}

\author{Daniele Quercia}
\orcid{0000-0001-9461-5804}
\affiliation{
  \institution{Nokia Bell Labs}
  \city{Cambridge}
  \country{United Kingdom}}
\affiliation{
  \institution{Politecnico di Torino}
  \city{Turin}
  \country{Italy}}
\email{quercia@cantab.net}

\renewcommand{\shortauthors}{Edyta Bogucka, Marios Constantinides, Sanja Šćepanović, and Daniele Quercia}

\begin{abstract}
Communicating the risks and benefits of AI is important for regulation and public understanding. Yet current methods such as technical reports often exclude people without technical expertise. Drawing on HCI research, we developed an Impact Assessment Card to present this information more clearly. We held three focus groups with a total of 12 participants who helped identify design requirements and create early versions of the card. We then tested a refined version in an online study with 235 participants, including AI developers, compliance experts, and members of the public selected to reflect the U.S. population by age, sex, and race. Participants used either the card or a full impact assessment report to write an email supporting or opposing a proposed AI system. The card led to faster task completion and higher-quality emails across all groups. We discuss how design choices can improve accessibility and support AI governance. Examples of cards are available at: \url{https://social-dynamics.net/ai-risks/impact-card/}.
\end{abstract}

\ccsdesc[500]{Human-centered computing~Interactive systems and tools}
\ccsdesc[500]{Human-centered computing~Collaborative and social computing}

\keywords{impact assessment, regulation, artificial intelligence, data visualization}

\received{July 2024}
\received[revised]{December 2024}
\received[accepted]{March 2025}

\begin{teaserfigure}
    \centering
    \includegraphics[width=0.33\textwidth]{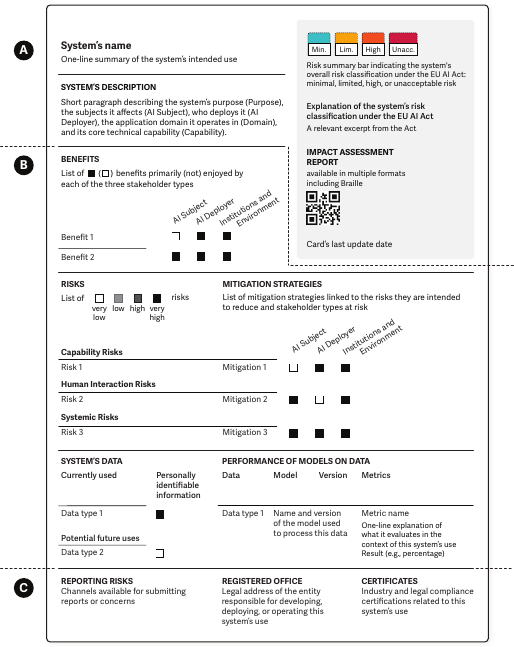}
    \caption{Template of the Impact Assessment Card showing AI risks and benefits in plain terms. Section A includes the system's name, purpose, and EU AI Act risk level. Section B outlines benefits, risks with mitigations, and technical details. Section C lists reporting channels, registered office, and certifications. Examples of cards for four AI systems appear in Appendix \ref{app:cards-after-updates}--\ref{app:card-digital-system} and online at: \url{https://social-dynamics.net/ai-risks/impact-card/}.}
    \Description{figure description}
    \label{fig:teaser}
\end{teaserfigure}

\maketitle

\section{Introduction}
The transformative potential of AI in society requires a thorough understanding of its risks and benefits~\cite{unesco2023ethicalImpactAssessment, kawakami2023sensing}, with policymakers advocating that providing the public with algorithmic advice will improve risk predictions, and, in turn, lead to better and fairer algorithmic decisions~\cite{green2021algorithmic, fogliato2021impact}. This need has led to the creation of fully-fledged impact assessment reports as a way of identifying and mitigating potential risks associated with AI systems, and communicating AI's potential benefits to individuals, society, and the environment~\cite{metcalf2021constructingImpacts}. Producing such reports requires an in-depth grasp of the AI system, from its initial ideation to its real-world deployment. This includes knowledge of the training data, the underlying algorithms, and the effects these systems might have on society and the environment. Moreover, it is essential to effectively share this knowledge with all parties involved, including legal entities and the general public, whose rights are often affected by the AI systems~\cite{prabhakaran2022human}. As AI governance continues to evolve, impact assessment reports are set to become a legal requirement. The EU AI Act, for example, requires detailed reports on the impact of high-risk AI uses on human rights, the environment, and the public interest~\cite{EUAIAct_2024}. These reports aim to increase transparency regarding AI functionalities, hold corporations accountable for the ethical and societal consequences of their AI systems, and allow ordinary individuals to comprehend the risks and benefits of AI uses to make informed decisions about their adoption.

However, a review of more than 300 AI auditing tools found that discovering harms within AI systems and effectively communicating these harms have received far less attention than evaluating the technical performance of those systems \cite{ojewale2024towards}. Current reports on AI impact assessments, often filled with technical jargon \cite{liang2024s}, are mainly aimed at experts and can alienate ordinary individuals impacted by AI's societal integration. This creates a barrier to wider understanding and participation in AI-related discussions. Therefore, it is crucial to explore new methods of communicating the risks and benefits of AI uses that are inclusive and understandable to everyone.  

Drawing from the HCI and CSCW literature, as we shall see in \S\ref{sec:related}, we aim to simplify and communicate complex concepts pertaining to AI uses for broader public consumption. For example, the use of clear language, icons, metaphors, and color coding can make complex AI information more accessible to ordinary individuals~\cite{goodman2018impact, grunert2014sustainability}. With that aim in mind, we made two main contributions:

\begin{enumerate}
    \item Through an iterative design process, we conducted three focus groups with 12 participants who identified design requirements for an impact assessment card, and designed a set of speculative cards. The design requirements were grouped into two categories: those related to the information (i.e., what the card should contain), and those related to the design (i.e., how the card should convey the information). By reviewing these speculative cards and soliciting feedback from the research team, we designed our impact assessment card (Figure \ref{fig:teaser}, \S\ref{sec:formative_study}). 
    \item We evaluated our card's effectiveness for conducting a real-world task (e.g., a compliance expert typically writes emails to the ethics committee, recommending implementation of an AI or advising against it), and compared it against a baseline impact assessment report in an online study with 235 participants across three cohorts: AI developers, compliance experts, and ordinary individuals who reflect US Census in terms of age, sex, and race (\S\ref{sec:evaluation}).\hspace{0.1em}
    We found a strong preference for the card across the three cohorts, with ordinary individuals expressing the highest favorability. Its user-friendly and accessible format not only allowed for faster reading times but also enabled participants to execute the task more efficiently, resulting in higher-quality emails.
\end{enumerate}

We conclude by discussing how impact assessment cards can help assess AI risks, communicate its benefits, and support AI governance. Additionally, we explore design opportunities and potential applications of the cards across various contexts (\S\ref{sec:discussion}).
\section{Related Work}
\label{sec:related}
Next, we surveyed diverse lines of literature that our work draws upon, and grouped them into two areas: \emph{a)} ways of communicating AI uses' risks and benefits to technical roles (\S\ref{subsec:comms_technical}); and \emph{b)} HCI and CSCW literature concerned with communicating multi-faceted and complex concepts to ordinary individuals (\S\ref{subsec:comms_ordinary}).

\subsection{Communicating Risks and Benefits of AI Uses to Technical Roles}
\label{subsec:comms_technical}

Numerous responsible AI artifacts---defined as processes, tools, documentation templates, and other resources designed to support the ethical creation and use of AI \cite{kawakami2024responsible}---have been developed for data and model evaluation. However, these artifacts tend to focus more on facilitating risk and benefit communication among technical roles such as developers and engineers \cite{legalCards2015, judgmentCall, inclusiveCards2023, wang2024farsight}. \citet{gebru2021datasheets} introduced ``datasheets for datasets'' for comprehensive dataset descriptions including key test features, test outcomes, and potential biases. Similarly, Bender et al.~\cite{bender2018data} proposed ``data statements'' for dataset demographic overviews. For standardized model information, Mitchell et al.~\cite{mitchell2019model} suggested ``model cards'', describing the uses, performance, biases, and limitations of machine learning models. However, a study of completion rates for Hugging Face’s model cards~\cite{liang2024s} showed that developers often prioritize completing information on training details, neglecting environmental impacts and evaluations.

Expanding on these artifacts, the prevalent method of communicating the risks and benefits of AI systems to technical roles is through impact assessment reports \cite{stahl2023systematicReview, iso2023ManagementSystem, IATemplate_2024, AIDesign_2024, iso2025SystemImpactAssessment}. Stahl et al.~\cite{stahl2023systematicReview} described the impact assessment process as a systematic approach to comprehend the potential positive or negative consequences of an AI system. This process typically entails detailing the AI system's intended use and benefits, evaluating risks, and formulating mitigation strategies \cite{IATemplate_2024}. For example, the Responsible AI impact assessment template~\cite{microsoft2022Assessment} includes system information, identified risks, mitigation measures, and an impact summary. The algorithmic impact assessment~\cite{equalAI2023nist} further delineates system information into system tasks and operational contexts and categorizes risks as either organizational or those stemming from third-party technologies. These elements are also relevant in specialized fields, such as the algorithmic impact assessment for AI in healthcare~\cite{ada_lovelance} and the societal impact assessment for computing researchers in industry~\cite{SocietalIATemplate2025}. Collectively, these serve as the state-of-the-art example reports for detailing and communicating the risks and benefits of AI systems.

However, the EU AI Act~\cite{EUAIAct_2024} mandates documenting impacts not at the dataset, model, or AI system level, but for a specific AI system's use, which can be detailed through five components~\cite{Golpayegani2023Risk}: purpose (the AI's intended goal), AI deployer (the entity managing the AI), AI subject (individuals or groups affected by the AI), capability (the AI's technological feature), and domain (the sector of AI use). To help communicate risks and benefits in this format,~\citet{Hupont2024} proposed ``use cards'' that list, among other information, the system's intended use, impacted stakeholders, and Sustainable Development Goals to be supported by the use~\cite{sdgs}.

\subsection{Communicating Multi-Faceted and Complex Concepts to Ordinary Individuals}
\label{subsec:comms_ordinary}

Communicating AI's risks and benefits to the general public is challenging; however, HCI and CSCW studies provide strategies to simplify these complex concepts for non-experts \cite{visWhatWorks2021}. Scientific sketchnotes by \citet{fernandez2019multimodal} combine notes and sketches to introduce complex scientific topics to laypeople. \citet{shen2020designing} redesigned confusion matrices for binary classification to improve non-experts' understanding of machine learning model performance. They found that by contextualizing terminologies and using flow charts to indicate data reading direction, significantly improved comprehension. Similarly, \citet{kehrer2012visualization} explored various techniques for visualizing multifaceted scientific data such as abstract representations, data aggregation, and the strategic use of texture and color. The addition of color, particularly red, has been shown to significantly increase perceived risk, a phenomenon observed across multiple cultures, even though cross-cultural studies are limited \cite{colorPsychology_2015}. Orange and yellow are the next most commonly used colors for marking risk after red, although people often find it difficult to distinguish which of the two conveys a higher level of risk when used together \cite{colorPsychology_2015}. Additionally, using prominent typography further enhances the memorability of risk warnings \cite{warningTypography_1990}. The length of an artifact (e.g., a card) has also been linked to the comprehension and perceived trustworthiness of an AI. When testing shorter and longer versions of their AI Model Cards among non-experts, \citet{bracamonte2023effectiveness} found that longer versions of the cards were considered less understandable and interpretable compared to a short version. However, they also found that the short version had a slightly negative effect on the perceived trustworthiness of the AI. 
Moreover, \citet{kawakami2024responsible} identified additional challenges in ensuring that Responsible AI artifacts such as ``datasheets for datasets'' \cite{gebru2021datasheets}, effectively serve non-technical stakeholders, including regulators and civil society organizations. These challenges include a misalignment between the technical details provided and the specific decision-making needs of these stakeholders, insufficient clarity in conveying the real-world implications of AI risks, and limited opportunities for stakeholders to evaluate the artifacts. These barriers highlight the need for resources that not only simplify complex AI concepts but also actively engage non-technical actors in the broader governance ecosystem \cite{everydayAlgorithmAuditing2021, UserDrivenAlgorithmAuditing_2022}. For example, the AI Failure Cards use comic strips to depict real-world AI failures and include questions that help non-technical stakeholders suggest mitigation strategies \cite{tang2024ai}. Atlas of AI Risks maps AI uses in an interactive dashboard, showing their risks, benefits, mitigation strategies, and, where relevant, related incidents \cite{riskAtlas2024}.

Metaphors are a key tool designers use to shape and influence user expectations effectively or communicate complex information \cite{AIMetaphors2025}, especially in human-AI collaboration scenarios~\cite{khadpe2020conceptual}. One such metaphor is the use of labels to highlight specific product or service attributes, aiding consumers in making informed choices. This practice is prevalent in sectors such as agriculture~\cite{gorton2021determines}, food~\cite{jones2019front}, and energy~\cite{stadelmann2018different}. For example, ``nutrition labels'' in the food industry offer a simplified and comprehensible way for consumers to understand a product's nutritional value. Similarly, an impact assessment card for AI systems should distill complex information into a format that helps ordinary individuals understand the risks and benefits of AI uses such as trade-offs between accuracy and fairness of models \cite{trustModel}. AI Nutrition Facts~\cite{AINutritionFacts} adopted the metaphor of ``nutrition labels'' to describe AI services, covering aspects like model type, data use, data retention, and human oversight. Similarly, Open Ethics Label~\cite{AIElectricityLabel} uses the metaphor of ``energy labels'' to disclose details about AI services, including data provenance, algorithms, and their types of reasoning.

While detailed information is often available on the back of food packaging (similar to how information about the risks and benefits of AI uses is presented in full reports), it can be overly complex for many consumers. This complexity mirrors the challenges end-users encounter with AI documentation. The use of icons \cite{privacyLabels2009, lindley-aiLegibility-2020, scharowski2023certification}, charts \cite{trustModel, vizConflictingInformation, vizComprehension}, metaphors \cite{privacyLabels2009, AIMetaphors2025}, and straightforward language \cite{languageVis} can render this information more accessible to a diverse audience~\cite{goodman2018impact, grunert2014sustainability}. \hspace{5em}
For example, using labels with absolute instead of relative rates and conveying probabilities with frequencies (e.g., ``3 out of 10'') instead of percentages (e.g., ``30\%'') improves understanding of risks in low-numeracy audiences \cite{visWhatWorks2021}. Deliberate design choices can help not only in conveying the risks and benefits of a product but also in enhancing trust in it~\cite{tonkin2015trust, trustModel, visWhatWorks2021}. 
\smallskip

\noindent\textbf{Research Gap.} In summary, previous research on communicating the risks and benefits of AI uses has mainly targeted technical audiences, relying primarily on detailed impact assessment reports. However, the fields of HCI and CSCW provide a rich repository of strategies that can be leveraged to create artifacts designed for a wider audience. Our work seeks to bridge this gap by designing and developing an impact assessment card aimed at communicating the risks and benefits of AI uses to both technical and non-technical roles.

\section{Author Positionality Statement}
Before presenting our impact assessment card, we clarify our positionality to enhance understanding of the methodology, study design, data interpretation, and analysis~\cite{DarwinHolmes2020_Positionality}. We are situated in a Western country in the 21\textsuperscript{st} century, contributing as authors who are predominantly engaged in research within academia and industry at a large technology company. We have contributed to the design, development, and implementation of tools supporting Responsible AI, including guidelines, toolkits, and documentation templates \cite{constantinides2023method, AIDesign_2024, IATemplate_2024}. Our team includes four members---two women and two men---from Southern and Eastern Europe, representing diverse ethnic and religious backgrounds. Our combined expertise covers Responsible AI, human-computer interaction (HCI), data visualization, artificial intelligence, and natural language processing. These experiences and backgrounds influenced our data interpretation, the way we incorporated participant feedback into the template’s design and development, and the choice of real-world tasks. We recognize the importance of expanding the perspectives presented in this paper and encourage future contributions from individuals with diverse backgrounds, especially those from outside academia and industry.
\section{Design the Impact Assessment Card}
\label{sec:formative_study}
To design the impact assessment card, we followed a two-step method that combined insights from literature with findings from design activities. First, we reviewed prior studies to identify 14 design patterns used to communicate the risks and benefits of AI uses, which provided a foundation for subsequent speculative design activities conducted in three focus groups (\S\ref{subsec:focus_groups}). Second, we iterated on the outcomes of the focus groups, which included 12 speculative card designs and 8 design requirements. We analyzed the card designs to obtain a preliminary version of our card and then addressed the design requirements to prepare its refined version for the user study (\S\ref{subsec:design_card}).

\subsection{Identify Design Patterns From Literature and Conduct Speculative Design Activities in Focus Groups}
\label{subsec:focus_groups}

\subsubsection{Identify Design Patterns From Literature}
We started by analyzing three systematic literature reviews that compile tools for communicating AI risks and benefits, as well as labels for trustworthy AI~\cite{cihon-ai-certification-2021, stuurman2021, ojewale2024towards}. We extracted an additional set of 4 design patterns from these papers such as nutrition labels for datasets~\cite{holland-nutritionLabel-2018}, icons for AI legibility~\cite{lindley-aiLegibility-2020} or certificates for machine learning methods~\cite{morik-careLabel-2022}. Finally, we reviewed studies in agriculture~\cite{gorton2021determines}, food~\cite{jones2019front}, and energy~\cite{stadelmann2018different}, in which labels have been effectively employed to communicate complex information, such as sustainability scores, to consumers. This review resulted in one additional design pattern. For the food and energy domains, we did not find any new patterns, as the food metaphor was already used in nutrition labels for datasets~\cite{holland-nutritionLabel-2018}, and the energy efficiency metaphor was used in the AI ethics label~\cite{stuurman2021}.\hspace{0.1em} 
The only new pattern came from the agricultural domain: a data hazard label inspired by chemical hazard labels, such as those for flammable substances~\cite{data_hazards}.

We grouped the 14 design patterns derived from the literature into two categories (Appendix~\ref{app:design-patterns}, Figure~\ref{fig:design-patterns}): visual representation (i.e., common visual elements for communicating AI uses' risks and benefits), and layout (i.e., how visual elements are combined together). For visual representation, we identified the use of textual descriptions, numeric values, links, tags, icons, charts, data samples, checkboxes, and metaphors (e.g., traffic lights). For the layout, we identified the use of lists, tables, rankings, grids, and groups. The list of the design patterns may not be exhaustive but rather served as a kickstarter in the speculative activities during the focus groups. To facilitate similar activities, we made the list available at: \url{https://social-dynamics.net/ai-risks/impact-card/}.

\aboverulesep=0ex
\belowrulesep=0ex 
\begin{table*}
\caption{Participant demographics of the three focus groups. GID: focus group identifier; PID: participant identifier; Role: AI developer ($R_{D}$), compliance expert ($R_C$), and ordinary individual ($R_O$).}
\label{tab:formative_participants}

\begin{tabular}{lllllll}
\hline
GID & PID & Age & Gender & Role & Institution & Location \\
\hline
\multirow{4}{*}{G1} & P1 & 29 & F & $R_C$ & Academia &  UK \\ 
& P2 & 25 & M & $R_{D}$ & Academia & UK  \\ 
& P3 & 34 & F & $R_O$ & Industry  & Germany   \\ 
& P4 & 28 & M & $R_{D}$ & Industry & UK   \\ 
\hline
\multirow{4}{*}{G2} & P5 & 26 & F  & $R_O$  & Academia & UK   \\
& P6 & 59 & M & $R_O$  & Industry & Belgium  \\ 
& P7 & 27 & M & $R_C$ & Academia & UK   \\ 
& P8 & 35 & F & $R_{D}$  & Industry & UK  \\ 
\hline
\multirow{4}{*}{G3} & P9 & 33 & M  & $R_{D}$  & Industry & UK  \\ 
& P10 & 26 & F & $R_O$ & Industry & Portugal  \\  
& P11 & 25 & F & $R_C$ & Academia & UK  \\ 
& P12 & 27 & M & $R_{D}$ & Academia & UK  \\ 
\hline
\end{tabular}
\end{table*}

\subsubsection{Identify Design Requirements and Design Speculative Cards in Three Focus Groups}
\mbox{}

\noindent\textbf{Participants.} We used snowball sampling and began by identifying initial participants (5) through an internal mailing list at our organization. These participants were asked to refer additional participants from their own networks, expanding the sample size through successive referrals. We recruited a total of 12 participants (6 female, 6 male, with a median age of 27.5 years old) representing three different cohorts: AI developers (5), compliance experts (3), and ordinary individuals (4). We then conducted 3 focus groups of 4 participants each, ensuring each group had at least one participant from each cohort (Table \ref{tab:formative_participants}).
\smallskip

\noindent\textbf{Procedure.} The focus group consisted of five phases: a \emph{warm-up} activity, \emph{briefing}, a \emph{brainstorming} task, a \emph{speculative design} task, and \emph{debriefing}. In the \emph{warm-up}, participants stated their name and answered, ``If you could be an AI, which one would you be, and why?'' to foster creativity. During the \emph{briefing}, participants were introduced to impact assessment cards through two examples: AI Nutrition Facts~\cite{AINutritionFacts} and Open Ethics Label~\cite{AIElectricityLabel}, which describe AI services as ``nutrition'' and ``energy'' labels. They then moved to a Figma~\cite{figma} to engage in the two tasks.

During the \emph{brainstorming task}, we aimed to surface the needs of different cohorts for the impact assessment card. It started with an idea generation session where participants used notes to brainstorm about their needs in terms of the card's functionality, specific tasks they think the card will assist with, information content, and format. This was followed by categorizing the ideas into four types of requirements: ``must have'', ``should have'', ``could have'', and ``won't have''. This categorization is based on the MoSCoW method for managing trade-offs during product design~\cite{achimugu2014systematic}. Must-have requirements describe critical features; should-have indicate important but not critical features; could-have describe desirable features (e.g., which could improve user experience); and, won't have indicate features that have been considered but explicitly decided against.

During the \emph{speculative design task}, we aimed to surface visual representations of the impact assessment cards that align with our participants' needs identified in the brainstorming task. Participants were first asked to read a report that documents the risks and benefits of a hypothetical AI system for identifying crime hotspots in public spaces using CCTV footage \cite{AIDesign_2024}. Informed by previous studies~\cite{liao2022, malak_vsd}, the use of a hypothetical system with real-world applicability served as a way to help participants contextualize their speculative designs. Participants were then introduced to the 14 design patterns derived from the initial literature review (Appendix~\ref{app:design-patterns}, Figure~\ref{fig:design-patterns}), and were given five minutes to review all the patterns. Finally, they were asked to create a speculative design for an impact assessment card for the hypothetical system. They could either build upon the existing patterns or propose new ones. Participants were instructed to sketch their design using pen and paper, photograph it, and upload it to the Figma board. At the end of this task, each participant explained their design choices.

The focus group ended with a \emph{debriefing} to summarize the main ideas that emerged and provided an opportunity for participants to share final recommendations for the card. Each session lasted 1 hour, and was video- and audio-recorded with participants' consent. The audio was automatically transcribed by the video conferencing software. The study was approved by our organization.
\smallskip 

\noindent\textbf{Analysis.} 
To derive design requirements, we conducted a qualitative analysis of the recordings and audio transcripts of the focus groups, which include participants' expressed ideas during the brainstorming and speculative design tasks and debriefing sessions. Two authors thematically analyzed these ideas following an inductive approach~\cite{saldana2015coding, miles1994qualitative, mcdonald2019reliability, braun2006thematic}. The authors used Figma~\cite{figma} to collaboratively create affinity diagrams based on these participants' inputs. Over the course of six meetings, totaling 16 hours, they discussed and resolved any disagreements that arose during the theme analysis process. From each resulting theme, the authors derived a design requirement and provide example(s) of how our participants phrased the requirement.
\smallskip

\begin{table*}[t!]
\centering
\caption{Eight design requirements identified during the focus groups, grouped into those on information and on design, along with implementation decisions for the preliminary (A) and refined (B) version of the card.} 
\scalebox{0.72}{
\begin{tabular}{p{2.45cm}|p{5.8cm}|p{4.75cm}|p{4.75cm}}
\hline
\textbf{Theme} & \textbf{Design Requirement} & 
\textbf{\makecell[l]{ Implementation Decision in A \\(Preliminary Version)}} & \textbf{\makecell[l]{Implementation Decision in B \\ (Refined Version)}} \\ \hline

\multicolumn{4}{l}{\textbf{Requirements on information}}  \\ \hline
Data & \textbf{R1.} Show information about the system's data, distinguish between its essential and non-essential, and personal identifiable and document later uses of the data & Add a table with icons and tags to distinguish between data types & Replace the table with the heatmap of data types \\ \hline

Model & \textbf{R2.} Show information about models and its performance & 
Add a table detailing model names, versions, and accuracies & Align the model table with the heatmap of data types \\ \hline

Benefits & \textbf{R3.} Show information about the system's benefits enjoyed by individuals and the environment influenced by the system & Add a list of benefits for direct stakeholders (AI deployers using the system and AI subjects affected by the system) and indirect (related institutions and environments) & Replace the list with the heatmap of stakeholders enjoying the benefits \\ \hline

\begin{tabular}[c]{@{}l@{}} Risks and\\ Mitigations \end{tabular}  & \textbf{R4.} Show information about system’s risks faced by individuals and the environment influenced by the system and potential mitigation strategies & Add a list of risks and a list of mitigations for direct and indirect stakeholders & Combine the two lists into a table with risks, mitigations and the heatmap of affected stakeholders \\ \hline

\begin{tabular}[c]{@{}l@{}} Reporting and \\ Governance \end{tabular} & \textbf{R5.} Show information about reporting mechanisms and who it's responsible for its governance & Add two sections for reporting mechanisms and compliance certifications & Combine sections and include the registered office address \\
\hline

\multicolumn{4}{l}{\textbf{Requirements on design}}  \\ \hline
\begin{tabular}[c]{@{}l@{}} Accessible \\ Communication \end{tabular}  & \textbf{R6.} Use accessible textual and visual communication for quick decision-making & Use concise language, avoid technical terms, add summary bar with the system's risk classification & Add concise description of the system including the direct stakeholders, refine the summary bar and provide its explanation \\ \hline

Accessible Medium & \textbf{R7.} Use medium that is accessible both physically and digitally even by people with different abilities and those visually impaired & Link the card with a QR code to a longer version of the impact assessment report, ensure print and Braille compatibility, use high-contrast design & Improve the contrast ratios in the summary bar \\ \hline

\begin{tabular}[c]{@{}l@{}} Cultural\\ Inclusivity \end{tabular} & \textbf{R8.} Use inclusive textual and visual communication for accommodating diverse cultural perspectives  & Avoid the use of culturally sensitive colors and icons & Remove the icons and tags for data types \\ 
\hline

\end{tabular}
}
\label{tab:design_requirements}
\end{table*}

\noindent\textbf{Results.} Our participants envisioned many potential uses for the card, spanning both everyday consumer scenarios and specialized professional contexts. These included comparing the quality of different AI-based services (2 mentions), understanding the safety of these services before deciding to purchase or subscribe to them, often under time pressure (4 mentions), and contacting relevant authorities or support teams for concerns when problems occur (4 mentions). To support these and similar uses, participants identified eight requirements for the card. We grouped them into two main categories (Table~\ref{tab:design_requirements}): those related to the information (i.e., what the card should contain)---\emph{R1-5}, and those related to the design (i.e., how the card should convey the information)---\emph{R6-8}.
Regarding the \emph{information}, we identified five requirements about the: \emph{data}, \emph{model}, \emph{benefits}, \emph{risks and mitigation strategies}, and \emph{governance and reporting}. Data is about ensuring that card users are informed about the types of data the system collects to enable its use. For example, P2, a developer, suggested that \emph{``the card should include what data an AI system accesses about a certain user, and how this data is used (i.e., is it used to train the model or is it stored and for how long)''}. P1, a compliance expert, saw this section of the card as a way to \emph{``help people to choose whether to provide their data for a system, as when signing up to a new service or purchasing tech (e.g., Alexa, Notion AI)''}. Model requirement is about making the inner workings of the system's models transparent to the users. P4, a developer, emphasized that the card \emph{``should specify all data sources that the models have been trained on, pass assessments, and report the models' accuracy''}. Benefits is about informing users about the broader effects of the system, including its positive impact on people and the planet. P9, a developer, explained they \emph{``want to see the actual value of the system based on the collected data''}.

\noindent Risks and mitigation strategies are centered on acknowledging and addressing the potential negative impacts associated with the system's operation. P11, a compliance expert, stated that the card should report \emph{``what is the risk-level of the AI system (and how this risk level is decided)''}. Similarly, P2, a developer, stated that \emph{``[the card] should assist potential users of AI systems to quickly understand how `safe' they are before deciding to purchase or subscribe to them with a star-based rating''.} Finally, governance and reporting is about the system's regulatory compliance, accountability mechanisms, and the availability of channels for reporting concerns or risks. For example, P3, an ordinary individual, stated that \emph{``[the card] should tell right away how safe the product is and who certified it''.}

Regarding the \emph{design}, we identified three requirements about: \emph{accessible communication}, \emph{accessible medium}, and \emph{cultural inclusivity}. Accessible communication is about ensuring that all system-related information is presented in a manner that is easily understandable and accessible to a wide range of users. For example, P1, a compliance expert, stated that \emph{``the card should use language that is understood by everyone''}, while P8 and P6 stated that it should be \emph{``simple and straightforward''} and \emph{``understandable at a glance''}, respectively. Accessible medium emphasizes the need for the system's information and reports to be accessible across various formats and platforms, catering to diverse user needs. For example, P3, an ordinary individual, pointed towards the idea of providing\hspace{2em}
\emph{``access to more information about the product (e.g., QR Code)''}. Similarly, P2, a developer, stated that the card \emph{``should be available in both physical and digital form, depending on the type of system. An AI-powered smart speaker should have the card printed on the box, but an online subscription-based AI system like ChatGPT should have a digital equivalent shown to the user right before they complete registration''.} Finally, cultural inclusivity involves designing the system to be considerate of and respect diverse cultural backgrounds and perspectives. For example, P5, an ordinary individual, expressed that \emph{``[the card] should be culturally sensitive (e.g. colors used to signify bad vs. good)''.}

\begin{figure*}[!t]
  \centering
\includegraphics[width=\textwidth]{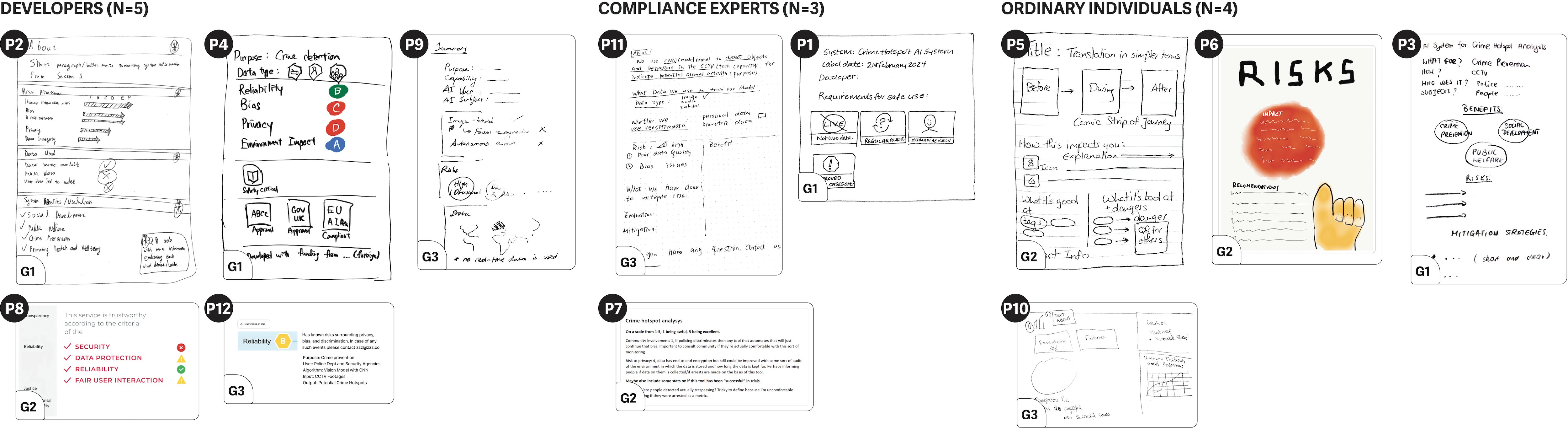}
  \caption{Speculative impact assessment cards created by 12 participants (P1-12) during three focus groups (G1-G3), sorted by cohort and layout. }
  \label{fig:design_patterns_and_cards}
\end{figure*}

\subsection{Iterate on the Results of the Activities to Design the Impact Assessment Card}
\label{subsec:design_card}

\subsubsection{Review the Cards From the Speculative Activities to Obtain the Preliminary Version of the Card}

\smallskip
The speculative design task during the focus groups resulted in a set of 12 speculative cards (Figure~\ref{fig:design_patterns_and_cards}) for the crime hotspot analysis system. To demonstrate the generalizability of our card across a variety of AI systems, we chose to implement the card for a similar system that processes personal image data and is more often encountered by ordinary individuals---a biometric supermarket checkout. To do so, we reviewed the set of cards designed by our participants, and for each design requirement (Table~\ref{tab:design_requirements}), we devised a set of implementation decisions that guided our initial card's design for the biometric checkout system.
\smallskip

\noindent\textbf{Implementation decisions for meeting requirements on information.}
To communicate system's data as per \emph{R1}, we introduced a two-column table inspired by the data nutrition labels \cite{holland-nutritionLabel-2018}, with one column for essential data (mandatory for system operation) and another for non-essential data (not critical for operation). Based on the speculative cards from P4, P5 and P8, each collected data is listed as a row with an icon indicating its format (e.g., image icon for image data). Based on P2's and P11's designs, each data is accompanied with tags indicating whether it contains personally identifiable information (as defined by GDPR) and whether it can be potentially re-used in other AI systems. To show model information as per \emph{R2}, we created a section documenting the performance of system models in accordance with the guidelines outlined for the model cards \cite{mitchell2019model} and card from P12. This section is also structured as a table, listing each collected data with corresponding columns for the model's name, version, and accuracy. While we primarily report on accuracy, the table can be extended to include other relevant metrics (e.g., error rates or confidence intervals). To communicate the benefits of the system's use as per \emph{R3}, we included a section listing these benefits (as suggested by P2, P3, P11) for three key stakeholders identified in the EU AI Act~\cite{EUAIAct_2024}: direct AI deployers (those using the system), AI subjects (those affected by the system)~\cite{Golpayegani2023Risk}, and indirect institutions and the environment. Using these same stakeholders as a reference, we structured the next section to follow \emph{R4} by pairing stakeholder-specific risks with potential mitigation strategies. \hspace{2em}
These pairings draw on the cards of P3, P6, P10, and P11. To facilitate reporting and governance as per \emph{R5}, we incorporated two sections: one providing information on reporting channels (e.g., dedicated email, phone number) and another showing compliance certifications (as seen on cards by P5, P11, P12) and a QR code linking to the full assessment report.
\smallskip

\noindent\textbf{Implementation decisions for meeting requirements on design.} 
To ensure accessible communication as per \emph{R6}, we refined the language to contain short phrases (maximum 50-65 characters or 8-11 words) and non-technical terms (as seen on cards by P2 and P3). This resulted in a Flesch-Kincaid Grade Level score of 11, indicating suitability for readers aged 16-17. Additionally, we introduced a summary bar similar to those found on food labels and drawn on cards by P4, P10, and P12, denoting the one-letter shortcuts for the system's overall risk classification as per the EU AI Act (with M for Minimal, L for Limited, H for High Risk, and U for Unacceptable risk) \cite{EUAIAct_2024}. To ensure accessible medium as per \emph{R7}, we linked the card with a QR code (as suggested on cards P2 and P5), allowing digital access to the full impact assessment report in print- and Braille-friendly formats. We further improved the card's readability by opting for a high-contrast design, with white background, ample white spaces, and black text in a 14-point sans-serif font with 125\% interline spacing to prevent text overcrowding (as in the medical leaflets \cite{InformationDesignLeaflet}). To ensure cultural inclusivity as per \emph{R8}, we refrained from employing culturally sensitive or strongly expressive colors and icons such as multiple shades of red for risk levels (visible on cards P4, P6). Instead, we selected a consistent color scheme for our risk summary bar based on established guidelines for the cross-cultural use of color in warnings \cite{colorPsychology_2015}: red for unacceptable uses, dark orange for high-risk uses, yellow for limited-risk uses, and blue for minimal-risk uses.

\begin{figure*}[t!]
  \centering
\includegraphics[width=0.95\textwidth]{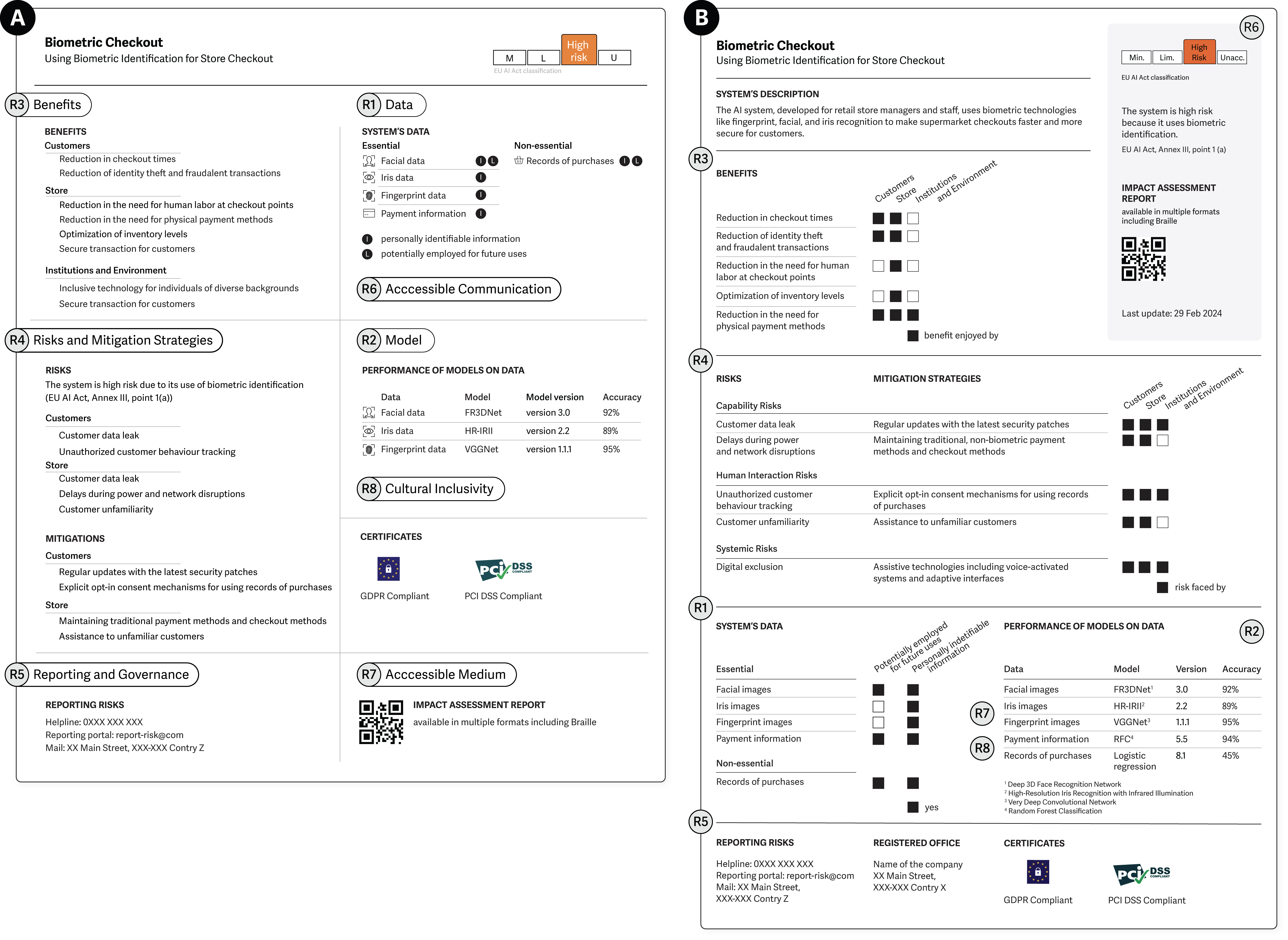}
  \caption{Impact assessment cards: preliminary (A) and refined (B) versions for the biometric checkout system. Both versions meet the 8 design requirements identified during the focus groups: present information on the system's data (R1), model (R2), benefits (R3), risks and mitigation strategies (R4), and governance (R5), while ensuring accessible communication (R6), medium accessibility (R7), and cultural inclusivity (R8). The refined versions were the result of four design iterations in the team. The final version of this card is provided in Appendix \ref{app:cards-after-updates} (Figure \ref{fig:updated-card-biometric-checkout}), together with the cards for three other systems: the license plate detector (Appendix \ref{app:cards-after-updates}, Figure \ref{fig:updated-card-license-plate-detector}), the music recommender system (Appendix \ref{app:card-digital-system}, Figure \ref{fig:card-music-recommender}), and the housing benefit allocation assistant (Appendix \ref{app:card-digital-system}, Figure \ref{fig:card-benefit-assistant}).}
  \label{fig:card_and_report}
\end{figure*}

Figure \ref{fig:card_and_report}A presents the first version of the card (nine sections). The top section contains the header with the AI system's name, its intended use, and a risk summary bar. The remaining sections are organized into two columns. The left column consists of four sections addressing various types of impact (benefits, risks, mitigation strategies) and providing information on reporting mechanisms. The right column contains technical details (system's data and model information), compliance certifications, and a QR code for accessing the full impact assessment report.

\subsubsection{Gather Recommendations From the Research Team on the Preliminary Version of the Card}
During the card's development, the first author conducted five sessions with the research team, progressively integrating feedback into refined versions of the card. By the time version 4 was completed, all feedback was implemented, and iterations ended with the following changes.
\smallskip

\noindent\textbf{Recommendations for meeting requirements on information.} To provide a clearer picture of the system's data (\emph{R1}), we transformed the two-column table into a heatmap. Essential and non-essential data is now displayed in a single column, with adjacent checkboxes replacing the icons and tags. This format enables easier recognition of patterns (e.g., excessive collection) and addition of new criteria (e.g., information about the source of data, licensing, real-time processing), without breaking the card's layout. To gain a better understanding of the model's effectiveness (\emph{R2}), we aligned the model performance section with the data one. Each row of the data's heatmap is linked to a specific model that uses the data and its overall performance. This integration simplifies the evaluation process. To improve the presentation of benefits (\emph{R3}), we explored alternative ways of grouping them. That is because we observed that the benefits were being repetitively listed across the AI deployer and AI subject---direct stakeholders. Similarly to the data, we introduced a heatmap with checkboxes for two key purposes: to clearly indicate the benefits that apply to each stakeholder, and to allow for potential expansion of the stakeholders' list. We also noted that, like benefits, risks were repetitively listed across different stakeholders. To better contextualize them as per \emph{R4}, we made three iterations. First, we categorized them according to capability, human interaction, and systemic risks, aligning with a framework for evaluating sociotechnical harms \cite{weidinger2023sociotechnical}. Next, for each risk category, we included a set of mitigation strategies. Finally, we used a heatmap to indicate the relevance of each risk to stakeholders, after considering the mitigation strategies. These iterations resulted in one section presenting a holistic view of risk management, enabling readers to see both the problem and the solution in one place. To improve the presentation of reporting and governance information (\emph{R5}), we restructured the section to combine risk reporting methods and certifications, while also expanding it to include details about the registered office (e.g., the official address of the legal entity responsible for the development and deployment of the system). This helps to build confidence in the system's transparency and adherence to legal standards, reinforcing readers' trust and assurance.
\smallskip

\noindent\textbf{Recommendations for meeting requirements on design.} To improve communication accessibility (\emph{R6}), we made two iterations: expanding the header and introducing a corner box. In the expanded header, we added a concise description of the system's core aspects using a five-component format \cite{Golpayegani2023Risk}: purpose, overseeing AI deployer, affected AI subject (direct), application domain, and technical capability enabling its use \cite{vocabularyCapabilities}. In the corner box, we added the risk summary bar, which we refined by replacing vague one-letter shortcuts with more informative abbreviations.\hspace{2em}
Below this bar, we provided explanations for each risk level (e.g., being high risk) and linked these to relevant articles from the EU AI Act \cite{EUAIAct_2024}. We also included a QR code for the full report and the date of the card's last edit.

We also refined the language describing the collected data to remove any ambiguities regarding the types of data collected. We iteratively transitioned from general terms in version 1 of the card (e.g., ``facial data'') to more precise descriptions in version 4 (e.g., ``facial images''). To improve medium accessibility (\emph{R7}), we revised the risk classification colors in the summary bar and improved their contrast ratios. Finally, to improve cultural inclusivity (\emph{R8}), we removed the icons representing the types of data collected. Although they work well for systems processing few datasets, their creation becomes problematic as the system expands to multiple datasets or more complex data types. Moreover, the use of numerous icons on a small card could lead to visual clutter, compromising the clarity of the information presented. 

\subsubsection{Refined Version of the Card}
Figure \ref{fig:card_and_report}B presents the refined version of the card. The top section of the card contains the expanded header and a corner box. The central section features the system's benefits, the risk management framework with combined risks and mitigation strategies, and the technical details on data and models. The bottom section contains information on reporting mechanisms, registered office and compliance certifications.
\section{Evaluate the Impact Assessment Card}
\label{sec:evaluation}

Having refined the card, we then evaluated it in a large-scale online study. The study's goal was to explore the effectiveness of the card to communicate the risks and benefits of AI uses in a way that is accessible beyond technical roles. Next, we describe our study's design (i.e., setup (\S\ref{subsec:ls_setup}), execution (\S\ref{subsec:ls_execution}), metrics (\S\ref{subsec:ls_metrics}), and results (\S\ref{subsec:ls_results})).

\begin{figure*}[t!]
  \centering
\includegraphics[width=\textwidth]{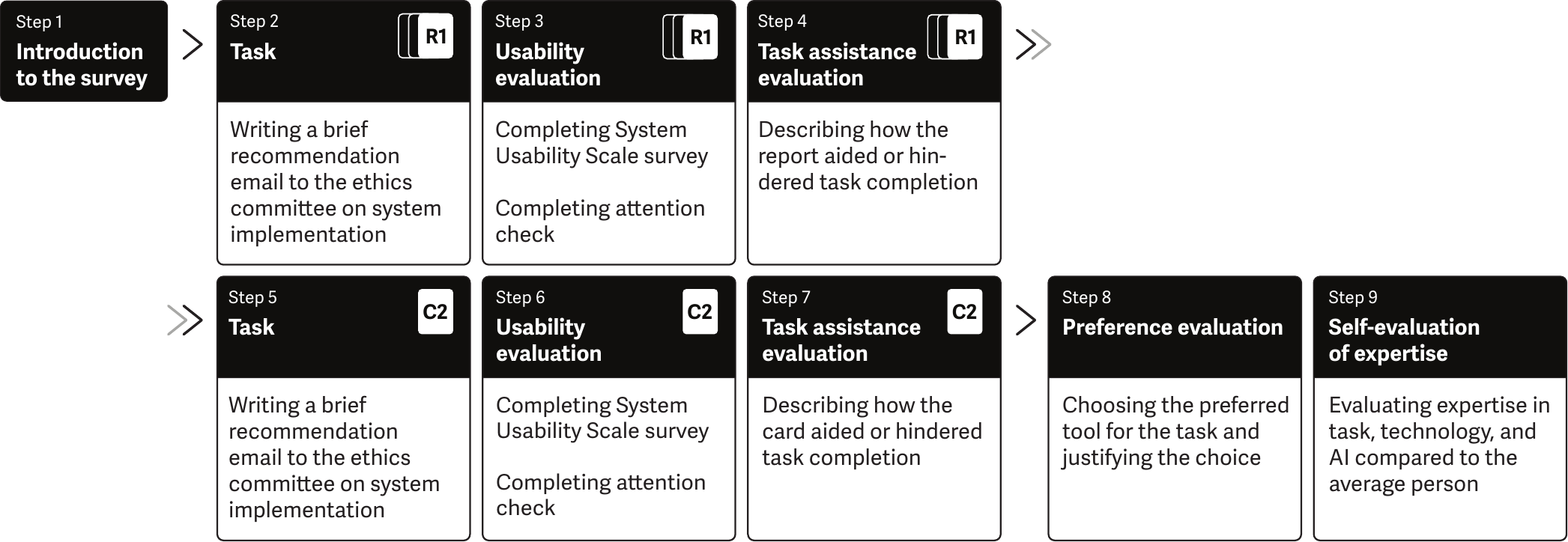}
  \caption{The online study involved 9 steps. Initially, participants received a brief introduction to the survey and tasks (Step 1). Then, they interacted with the first randomly assigned treatment (e.g., R1 - a report for the biometric checkout), completing a task (Step 2). Subsequently, they assessed the usability (Step 3) and assistance (Step 4) of the treatment. This process was repeated for a second treatment (Steps 5-7) depicting a different AI system (e.g., C2 - a card for the license plate detector). Finally, participants selected their preferred treatment for the task (Step 8), and self-evaluated their knowledge about the task, technology, and AI.}
  \label{fig:survey_setup}
\end{figure*}

\subsection{Setup}
\label{subsec:ls_setup}
We developed a web-based survey that included a real-world task to be performed either with the card or with the impact assessment report as baseline (Figure~\ref{fig:survey_setup}, Step 2 and Step 5).

\smallskip
\noindent\textbf{Task.} We defined a task related to the AI system that participants from each cohort might typically perform as part of their jobs~\cite{matcha} or interactions with AI: writing recommendation and feedback emails. This task was formulated based on insights from three areas: our focus groups about practical actions people take in response to AI systems affecting their lives, including the frequent need to contact relevant authorities or support teams when problems occur; conversations with AI practitioners and compliance experts in our organization about tasks in AI approval processes~\cite{ojewale2024towards}; and previous user studies on writing AI recommendations by different stakeholders~\cite{judgmentCall, madaioScopingStudy}.

Specifically, for an AI developer, the task was to read the card, and write a brief email to the ethics committee, recommending the implementation of the AI system or advise against it, in either case stating appropriate technical reasons. For a compliance expert, the task was to write an email to the ethics committee, recommending implementation of the system or advising against it. For an ordinary individual, the task was to write an email to the deployers who put in the AI system, asking them to take it out or thanking them, and in either case tell them why. The decision to reject or recommend the system was left entirely up to the participants based on their own judgment.

This task links the information from the card to three advanced decision-making skills typically supported by visualizations~\cite{vizTaskLevels}: problem-solving (determining appropriate actions), critical thinking (assessing and integrating information on risks, mitigation strategies, and benefits), and reasoning (forming logical arguments to justify actions). It leverages the specific skills and knowledge areas pertinent to each cohort: AI developers use their technical expertise, compliance experts apply their regulatory knowledge, and ordinary individuals draw from their user experience. We requested that emails from each cohort contain between 50 and 250 words, reflecting the typical length of descriptions used in model documentation~\cite{liang2024s}. This range ensures conciseness and sufficient detail for thematic analysis while minimizing participant fatigue.
\smallskip

\noindent\textbf{Control and Treatment for 2 AI systems.} In addition to the card (Appendix~\ref{app:cards-user-study}, Figures~\ref{fig:card-biometric-checkout}-\ref{fig:card-license-plate-detector}), we included a baseline condition to compare the card against (Appendix \ref{app:reports-user-study}, Figures \ref{fig:report-biometric-checkout}-\ref{fig:report-license-plate-detector}). 
We created an impact assessment report based on current state-of-the-art practices for communicating the risks and benefits of AI systems \cite{stahl2023systematicReview, equalAI2023nist, ada_lovelance}, drawing on examples from published reports \cite{microsoft2022Assessment, sherman2023riskProfiles, riskReports_2024}. These reports are issued by deployers of high-risk AI systems, as required under the EU AI Act \cite{EUAIAct_2024}, or by organizations seeking certification under AI management standards \cite{iso2023ManagementSystem}. The intended audience will primarily include market surveillance authorities, affected stakeholders, independent experts, and civil society organizations to ensure transparency and accountability. Our report mirrored the card's content (e.g., the system's use, data, models, evaluation, risks, mitigations, benefits, contact information, and certificates) but was more descriptive. We alternated between the card and report to eliminate any learning effects. Therefore, a participant was asked to execute the same task with the two conditions. To eliminate any effects from the type of AI systems shown in the card or report, we selected two hypothetical real-world AI systems that differ in risk level but are likely familiar to most participants. Next, we provided a brief description of each AI system. 

\begin{description}
    \item[Biometric Checkout.] This AI system uses biometric technology such as facial recognition to identify customers during the checkout process in a supermarket. By linking biometric data with payment methods and shopping histories, it enables seamless, secure checkout without physical cards or cash. The EU AI Act classifies this system as high risk \cite{EUAIAct_2024} due to its extensive use of biometric identification (Appendix~\ref{app:cards-user-study}, Figure \ref{fig:card-biometric-checkout}; Appendix~\ref{app:reports-user-study}, Figure \ref{fig:report-biometric-checkout}).
    \item[License Plate Detector.] This AI system uses cameras and image recognition technology to detect and read license plates of vehicles entering and exiting a supermarket car park. It can be used to monitor parking occupancy, enforce parking time limits, and ensure the security of the parking area. It is categorized as limited risk under the EU AI Act \cite{EUAIAct_2024} due to its processing of personally identifiable data (Appendix \ref{app:cards-user-study}, Figure \ref{fig:card-license-plate-detector}; Appendix \ref{app:reports-user-study}, Figure \ref{fig:report-license-plate-detector}).
\end{description}

\noindent Both systems, while beneficial to customers and stores, are considered risky under the EU AI Act \cite{EUAIAct_2024} due to real-time processing of personally identifiable information. Additionally, their excessive information collection and multi-model architecture enable potential future applications beyond their initially stated purpose. 

\subsection{Metrics}
\label{subsec:ls_metrics}
Independent to each cohort, we defined five metrics to capture the effectiveness in conducting the task. The first metric, \emph{task quality}, captured whether the resulting email was considered high quality. The email's quality was scored on a 5-point Likert scale based on how effectively the person used the information from the card or report to justify a recommendation for adopting or rejecting the system. An email scoring 1 was vague, applicable to any AI system, lacked a decisive call to action, and contained no arguments. An email scoring 5 was specific to the system described in the card or report, included a clear recommendation or rejection, and presented diverse arguments covering aspects such as risks, data, benefits, and mitigations. The second metric captured the factors \emph{influencing task quality} (both positively and negatively), with two open-ended questions: \emph{``In what ways did the card (or report) succeed to assist you in completing the task?''} and \emph{``In what ways did the card (or report) fall short to assist you in completing the task?''}. The third metric captured \emph{efficiency} in conducting the task, measured as the average time needed to read the card or report and complete the task. The fourth metric captured the \emph{usability} of the card or report, measured using the System Usability Scale~\cite{brooke1996sus}. Finally, the fifth metric captured the overall \emph{preference} for using the card or report for the task.

\subsection{Execution}
\label{subsec:ls_execution}
We recruited participants from Prolific~\cite{prolific} and surveyed them across three cohorts: \emph{a)} AI developers; \emph{b)} compliance experts; and \emph{c)} ordinary individuals (Table \ref{tab:large_scale_participants}). To recruit a sufficiently large number of participants for each cohort, we controlled for the participants' roles in the organization, the frequency of AI use in their jobs, and their geographic location using Prolific's built-in screeners. Additionally, we controlled for their expertise in the task at hand, technology in general, and AI through a self-reported assessment.

To recruit AI developers, we searched for participants who likely contribute to developing AI systems as part of their software engineering roles, using AI every day. We recruited 65 developers with a median age of 33 years: 7 female and 58 male, mostly White (54\%) and Asian (25\%). These participants were the most knowledgeable in technology and AI across the three cohorts. 

To recruit compliance experts, we searched for participants likely involved in revising AI systems as part of their legal roles, using AI at least 1-6 times a week. We recruited 65 experts with a median age of 42 years: 31 female and 34 male, mostly White (57\%) and Black (17\%). These participants were the most knowledgeable about the task at hand across the three cohorts, more knowledgeable in technology and AI than ordinary individuals, yet less so than AI developers. 

To recruit ordinary individuals, we used stratified random sampling to match US Census demographics \cite{census_ethnicity_2020, census_age_gender_2022} in terms of age (20\% in range 18-29 years, 17\% in range 30-39 years, 17\% in range 40-49 years,  16\% in range 50-59 years, 30\% over 60 years), sex (50\% female, 50\% male), and race (60\% White, 11\% Black, 10\% Mixed, 6\% Asian, 1\% Native American or Alaskan Native, 8\% Other)\footnote{Our research does not separately account for ordinary individuals who identify as Hispanic or Latino—the second-largest racial and ethnic group in the U.S.—because our recruitment followed the guidelines of the U.S. Census Bureau \cite{census_ethnicity_2020} and the U.S. Office of Management and Budget \cite{census_race_1997}. These sources define race using five categories---White, Black or African American, American Indian or Alaska Native, Asian, and Native Hawaiian or Other Pacific Islander, as reported above---while classifying Hispanic or Latino origin as an ethnicity. As a result, ordinary individuals in our sample who identify as Hispanic or Latino are recorded within these five racial categories.}. Compared to AI developers and compliance experts, as expected, ordinary individuals used AI less frequently in their jobs and had the least knowledge about the task at hand, technology, and AI. We restricted our participant pool to individuals living in the US for one main reason. Involving native English speakers ensured a clear understanding of the study materials, which strengthened the reliability of the findings. All participants were paid on average about \$12 (USD) per hour.
\smallskip

\begin{table*}[t!]
\centering
\caption{Self-reported knowledge and demographic characteristics of participants in a large-scale study.}
\label{tab:large_scale_participants}
\scalebox{0.9}{
\begin{tabular}{p{1.75cm}|p{3.45cm}|p{2.1cm}|p{2cm}|p{2cm}|p{1.5cm}p{1.2cm}}
\Xcline{1-6}{0.5pt}
\textbf{Control} & \textbf{Characteristic} & \textbf{AI developers} (n=65) & \textbf{Compliance experts} (n=65) & \textbf{Ordinary individuals} (n=105) & \textbf{US Census} \cite{census_ethnicity_2020, census_age_gender_2022} &  \\ \Xcline{1-6}{0.6pt}
                            & Task & 3.38 / 5 & 3.49 / 5 & 3.07 / 5 & - & \\ \cline{2-6}
                            & Technology in general & 4.20 / 5 & 3.60 / 5 & 3.30 / 5 & - & \\ \cline{2-6}
\multirow{-3}{*}{Expertise} & Artificial Intelligence & 3.82 / 5 & 3.32 / 5 & 2.96 / 5 &-  & \\ \Xcline{1-6}{0.6pt}
                            & 18-29 years & 30\% & 12\% & 20\% & 20\% &  \\ \cline{2-6}
                            & 30-39 years & 37\% & 23\% & 17\% & 18\% &  \\ \cline{2-6}
                            & 40-49 years & 18\% & 22\% & 17\% & 16\% &  \\ \cline{2-6}
                            & 50-59 years & 7\% & 30\% & 16\% & 16\% &  \\ \cline{2-6}
\multirow{-5}{*}{Age}       & 60 years and above & 8\% & 13\% & 30\% & 30\% &  \\ \Xcline{1-6}{0.5pt}
                            & Female & 11\% & 48\% & 50\% & 50\% &  \\ \cline{2-6}
\multirow{-2}{*}{Sex}       & Male & 89\% & 52\% & 50\% & 50\% &  \\ \Xcline{1-6}{0.5pt}                           
                            & White & 54\% & 57\% & 60\% & 62\% & \\ \cline{2-6}
                            & Black & 14\% & 17\% & 11\% & 12\% &  \\ \cline{2-6}
                            & Asian & 25\% & 15\% & 6\% & 6\% &  \\ \cline{2-6}
                            & Mixed  & 5\% & 9\% & 10\% & 10\% &  \\ \cline{2-6}
                            & \makecell[l]{Native American \\ or Alaskan Native} & 0\% & 0\% & 1\% & 1\% &  \\ \cline{2-6}
                            & Other & 2\% & 2\% & 8\% & 9\% &  \\ \cline{2-6}
\multirow{-7}{*}{Race} & Not specified & 0\% & 0\% & 4\% & - &  \\ \Xcline{1-6}{0.5pt}
\end{tabular}
}
\end{table*}

\noindent\textbf{Procedure.} We administered the survey on Prolific~\cite{prolific}. The survey first provided a brief introduction to the tasks, followed by the first task in which participants had to read either the card or report, and write the email, self-choosing to recommend or reject the system. This was followed by a series of questions to capture the usability of either the card or report, and questions about how they succeeded or fell short in assisting participants in completing the task. Participants repeated the same procedure for the second task. At the end, they indicated their overall preference for the card or report in completing the task.
To ensure response quality, we conducted two attention checks during the survey and implemented two survey design features. First, after reading the task instructions, participants encountered one of the attention-check sentences: \emph{``When asked for your favorite color, you must select `Blue'''} and \emph{``When asked for your favorite city, you must select `Rome'}''. Participants had to answer these checks correctly after each task. Second, we disabled pasting from external sources and editing previous responses to ensure original and thoughtful answers.

To control for the extent to which the answers depended on the participants' level of knowledge, we asked them whether they consider themselves more skilled or knowledgeable than most people for the task at hand, as well as for the technology in general and AI. This expertise was assessed using a 5-point Likert scale, ranging from ``Strongly disagree'' (1) to ``Strongly agree'' (5). 

\smallskip
\noindent\textbf{Analysis.} We performed both quantitative and qualitative analyses. For the quantitative analysis, we measured for both the task completed with the card and the task completed with the report: the average quality of the task; the average time to complete the task; the average SUS usability scores; and the percentage of participants who preferred the card or report for the task. \hspace{7em}
The evaluation of task quality was conducted by two contributing authors with expertise in Responsible AI, excluding the first author. Their assessment focused on whether the resulting emails were of high quality. Each email was rated by both authors on a 5-point Likert scale, ranging from poor (1) to excellent (5), based on five criteria: context, recommendation, risks, mitigations, and content clarity. To ensure consistency and accuracy in evaluations, the authors followed a predefined rubric (Appendix \ref{app:evaluation-rubric}). The rating process was blind to the experimental condition—the authors did not know whether an email was generated using the card or report. They were, however, aware of the cohort (developers, compliance experts, or ordinary individuals) since task formulation differed slightly across groups. The authors' assessments were largely consistent, with an inter-rater agreement of 85\%. In cases where the authors assigned different ratings, they discussed discrepancies in two assessment review meetings with the broader research team to reach a final decision.

We hypothesized five factors that might influence the quality of the task: the type of task (reject or recommend the system), the system (biometric checkout or license plate detector), the participant cohort (AI developers, compliance experts, or ordinary individuals), the participants' level of expertise (low or high), and, crucially, the treatment (card or report). We then conducted linear regression analyses and mean difference testing on these factors.

For the qualitative analysis, we thematically analyzed open-ended responses \cite{saldana2015coding, miles1994qualitative, mcdonald2019reliability, braun2006thematic} to understand the factors affecting task quality and preferences for using the card or report.

\subsection{Results}
\label{subsec:ls_results}
We received a total of 235 responses: 65 each from AI developers and compliance experts, and 105 from ordinary individuals. Next, we discuss the quantitative results based on our five metrics (\S\ref{subsubsec:quantitative}), followed up by qualitative results (\S\ref{subsubsec:qualitative}).

\subsubsection{Quantitative Results}
\label{subsubsec:quantitative}

\begin{figure}[t]
  \centering
  \includegraphics[width=\textwidth]{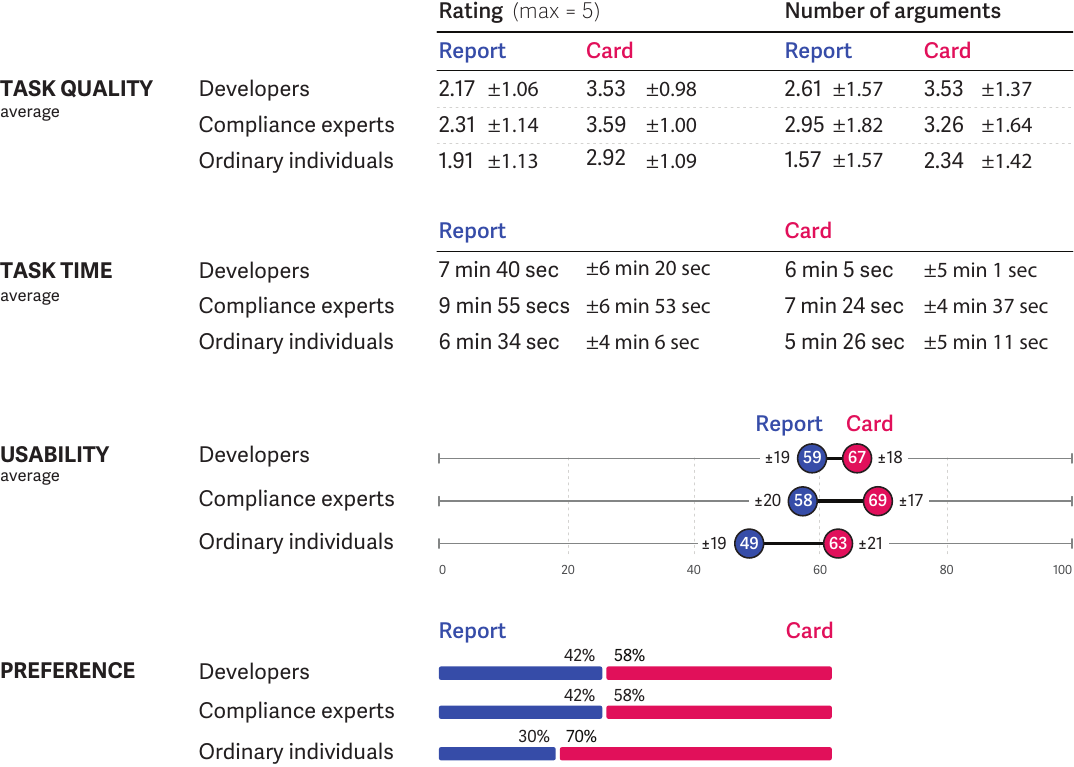}
  \caption{\textbf{Card outperformed report across all quantitative metrics and cohorts}. It helped produce higher quality emails in less time, while being more usable and preferred for the task.}
  \label{fig:results}
\end{figure}

\noindent\textbf{Regardless of the cohort, participants, on average, spent less time completing their tasks with the card than with the report with even better quality.}
Compliance experts achieved the highest average email ratings when using the card (3.59), followed closely by developers (3.53), and then ordinary individuals (2.92) (Figure \ref{fig:results}). They also took the longest to complete their tasks with the card (7 min 24 secs) (Appendix~\ref{app:other-regressions}, Table \ref{tab:regression-completion}).
\smallskip

\noindent\textbf{The card was rated higher in usability than the report, especially among ordinary individuals.}
Developers rated the card with an average SUS score of 67, compared to the report's score of 59, indicating a preference for the card's usability (Figure \ref{fig:results}) and generally positive user experience~\cite{sauro2011practical}. Compliance experts shared this view, scoring the card at 69, with the report at 58. However, this distinction was most pronounced among ordinary individuals, who gave the card a SUS score of 63, compared to a score of 49 for the report (Appendix~\ref{app:other-regressions}, Table \ref{tab:regression-usability}).
\smallskip

\noindent\textbf{All cohorts preferred the card over the report to execute the task at hand, with a higher preference among ordinary individuals compared to developers and compliance experts.} 
Over half of both developers and compliance experts, at 58\%, favored the card over the report (Figure \ref{fig:results}, Appendix~\ref{app:other-regressions}, Table \ref{tab:regression-preference}). In contrast, 70\% of ordinary individuals strongly preferred the card, compared to 30\% favoring the report.
\smallskip

\noindent\textbf{The most significant difference in the task quality was attributed to the use of card or report.} 
The most significant difference in task quality was due to the treatment (Table \ref{tab:quality-modelling-mixed-methods}, Table \ref{tab:sensitivity}), with the card receiving consistently higher ratings for task quality compared to the report. The type of task (advising for or against either of the two systems) and the participants' expertise levels did not impact the quality.

\aboverulesep=0ex
\belowrulesep=0ex 
\begin{table*}[ht]
\flushleft
\small
\caption{\textbf{The results of a linear mixed-effects regression analysis with task quality as the dependent variable. The most significant difference in task quality arises from the choice of treatment.} The coefficients represent the effect sizes for each factor relative to its reference category, with statistical significance indicated by: ** for $p<0.01$, and *** for $p<0.001$. Non-significant factors (p > 0.05) are also reported for completeness. Random effects were included to account for variability in task quality based on participants’ self-selected decisions to reject or recommend the system, ensuring fair comparisons across all fixed factors.}
\label{tab:quality-modelling-mixed-methods}
\scalebox{0.87}{
\begin{tabular}{p{3.1cm}|p{6.4cm}|p{1.95cm}|p{2.95cm}}
\toprule
\textbf{Factor} & \textbf{Values} & \textbf{Coefficient} & \textbf{$p$-value} \\
\midrule
Intercept &  & 2.795 & 0.000 \\ \hline
\multicolumn{4}{l}{\textbf{Type of task}}  \\ \hline
Recommendation & Reject \emph{vs.} Recommend & 0.880 & 0.325 \\
System & Plate Detector \emph{vs.} Checkout & 0.150 & 0.079 \\ \hline
\multicolumn{4}{l}{\textbf{Participant’s cohort}}  \\ \hline
Cohort & Developers \emph{vs.} Ordinary individuals & 0.131 & 0.257 \\
\cellcolor{highlight}\textbf{Cohort} & \cellcolor{highlight}\textbf{Compliance experts \emph{vs.} Ordinary individuals} & \cellcolor{highlight}\textbf{0.287} & \cellcolor{highlight}\textbf{0.007**} \\ \hline
\multicolumn{4}{l}{\textbf{Expertise levels}}  \\ \hline
Task Expertise & Low \emph{vs.} High & -0.013 & 0.819 \\
Technological Expertise & Low \emph{vs.} High & 0.055 & 0.392 \\
AI Expertise & Low \emph{vs.} High & -0.056 & 0.382 \\ \hline
\multicolumn{4}{l}{\textbf{Treatment}}  \\ \hline
\cellcolor{highlight}\textbf{Treatment type} & \cellcolor{highlight}\textbf{Card \emph{vs.} Report} & \cellcolor{highlight}\textbf{-0.987} & \cellcolor{highlight}\textbf{0.000***} \\
\bottomrule
\end{tabular}}
\end{table*}

\aboverulesep=0ex
\belowrulesep=0ex 
\begin{table*}
\caption{\textbf{The mean difference testing underscores the strong influence of treatment choice on task quality.} We conducted statistical significance testing on the mean differences between two factor values, presenting Mann-Whitney test p-values with the notations: * for $p<0.05$, ** for $p<0.01$, and *** for $p<0.001$.}
\label{tab:sensitivity}
\scalebox{0.8}{
\begin{tabular}{p{3.4cm}|p{7cm}|p{2.15cm}|p{1.6cm}|p{1.25cm}}
\hline
\textbf{Factor} & \textbf{Value Pair} & \textbf{Averages} & \textbf{Difference} &  \textbf{$p$-value} \\
\midrule
\multicolumn{5}{l}{\textbf{Type of task}}  \\ \hline
        Recommendation &                            Reject \emph{vs.} Recommend &        3.0 \emph{vs.} 3.014 &         -0.014 &            0.719 \\
                System &                     Plate Detector \emph{vs.} Checkout   &      2.801 \emph{vs.} 2.645  &         -0.156 &            0.139 \\ \hline
\multicolumn{5}{l}{\textbf{Participant's cohort}}  \\ \hline
Cohort &                    Developers \emph{vs.} Compliance experts &           2.852 \emph{vs.} 2.95 &         -0.098 &            0.534 \\
       \cellcolor{highlight}\textbf{Cohort} &    \cellcolor{highlight}\textbf{Developers \emph{vs.} Ordinary individuals} & \cellcolor{highlight}\textbf{2.852 \emph{vs.} 2.505} & \cellcolor{highlight}\textbf{0.347} &  \cellcolor{highlight}\textbf{0.011*} \\
       \cellcolor{highlight}\textbf{Cohort} & \cellcolor{highlight}\textbf{Compliance experts \emph{vs.} Ordinary individuals} &  \cellcolor{highlight}\textbf{2.95 \emph{vs.} 2.505} & \cellcolor{highlight}\textbf{0.445} & \cellcolor{highlight}\textbf{0.002**} \\ \hline
\multicolumn{5}{l}{\textbf{Expertise levels}}  \\ \hline
        Task Expertise &                                    Low \emph{vs.} High &       2.73 \emph{vs.} 2.714 &          0.015 &            0.852 \\
  Technological Expertise &                                    Low \emph{vs.} High &    2.691 \emph{vs.} 2.851 &          -0.16 &             0.25 \\
          AI Expertise &                                    Low \emph{vs.} High &       2.654 \emph{vs.} 2.803 &         -0.149 &            0.204 \\ \hline

\multicolumn{5}{l}{\textbf{Treatment}}  \\ \hline
\cellcolor{highlight}\textbf{Treatment type} &  \cellcolor{highlight}\textbf{Card \emph{vs.} Report} &  \cellcolor{highlight}\textbf{3.327 \emph{vs.} 2.12} & \cellcolor{highlight}\textbf{1.207} &  \cellcolor{highlight}\textbf{0.0***} \\
\hline
\end{tabular}
}
\end{table*}

\bigskip
\subsubsection{Qualitative Results}
\label{subsubsec:qualitative}
Through thematic analysis of participants' free-form answers, we identified key factors affecting their experience with the card and report, their overall preferences, and suggestions for improving the card. Participant quotes are referenced using $CP_N$, corresponding to their anonymized Prolific ID.
\smallskip

\noindent\textbf{The card was favored for its clear, concise presentation, and quick comprehension of the risks and benefits of AI uses, though some found it overly simplistic for complex decisions.}
On the positive side, the card was favored for its concise and straightforward presentation of information. Participants found it easier to digest, with visual elements and organized sections that allowed for quick understanding of the main risks and benefits of the presented AI systems. For example, CP9 stated that \emph{``[the card] assisted me by highlighting the risks, accuracy, and benefits,''} while CP23 appreciated \emph{``the card's structured overview of the system's components, facilitating the identification of key technical aspects of the AI system''}. A compliance expert, CP88, mentioned that \emph{``[the card] was easy to use and conveyed the gist of the AI system''}. Additionally, participants commented on the card's format to be \emph{``readily accessible to refer back to''} (CP129, a compliance expert). Participants also echoed the sentiment that despite spending less time with the card, it even helped them produce emails of higher quality. CP190, an ordinary individual, commented that \emph{``the best thing really is just that more thought went into making the card format more digestible and less intimidating, so that it would be easy to get what you need by reading it, without needing time to consult with more technical people to be sure you understand its material correctly''}. On the negative side, some participants noted that the card lacked the depth and detail typically found in the report. There were also mentions of the card being too simplistic for complex decision-making. CP6, a developer, felt that it \emph{``was a little simple, so I can't help but think there may be something missing in the big picture''}. Despite its concise format, some participants found the card too brief. For example, CP9, a developer, commented that \emph{``the card was brief which I enjoyed, however, it probably could have used a little more substance''}.
\smallskip

\noindent\textbf{The report was valued for its depth and details, though its complexity and dense format challenged quick comprehension and accessibility.}
On the positive side, participants appreciated the report for its detailed and comprehensive information, which helped them understand the AI system better. They mentioned that the report laid out the positive and negative aspects effectively and clearly in a well-structured way, providing a good foundation of knowledge. \hspace{7em} CP12, a developer, mentioned that it \emph{``helped explain why the system should be implemented, was organized and listed many different positive aspects''}. Similarly, CP152, an ordinary individual, mentioned that \emph{``the report succeeded in assisting me in completing the task by providing a wide array of information through which I could make a decision''}. On the negative side, a common critique was the report's complexity and length, making it difficult to quickly extract necessary information. Participants found it too detailed at times, with some sections containing excessive technical jargon. CP142, an ordinary individual, stated that \emph{``the report was overly wordy. It had lots of irrelevant information and technical jargon (e.g., the datasets that the models it uses were trained on).''} Similarly, CP157, another ordinary individual, stated that \emph{``the report felt very wordy. Reading it felt like I needed higher education to fully understand some aspects of the technology. I'm not sure if every day people would fully comprehend all the ins and outs of it''}. CP9, a developer, noted that \emph{``the report didn't advise in any direction''}. Additionally, some participants called out the lack of visual elements, which impacted their ease of understanding. CP75, a compliance expert, noted that \emph{``the report was too cluttered''}. Similarly, CP101, an ordinary individual, commented that \emph{``there was a lot of information to read and some of the reading I didn't understand had to read at least twice''.}
\smallskip

\noindent\textbf{Overall preference.}
The preference varied among participants. Some preferred the report for its thoroughness and detail, which they found necessary for making informed decisions. Others favored the card for its efficiency and simplicity, allowing for quicker comprehension and easier reference during their tasks. CP9, a developer, found the report more useful, stating that \emph{``the report had much more information that I could use to craft the email''}. Similarly, CP78, a compliance expert, stated a preference towards \emph{``the report as it provided a more detailed information about the AI system, impacts, risks and mitigation strategies enabling a thorough analysis and recommendation''}. Conversely, CP152, an ordinary individual, mentioned that \emph{``I prefer the card more than the report, because the card was more concise and clear in the information that it presented'' }. Similarly, CP101, a compliance expert, preferred the card because \emph{``it seemed easy to read and understand. It showed the entire plan like a minimalistic picture''}.
\smallskip

\noindent\textbf{Card improvements.} Participants also suggested ways to further improve the card by providing contextual information, reducing ambiguity, and enhancing its visual elements. Some participants found the cards too simplistic and lacking depth for full understanding. For example, CP97, a compliance expert, mentioned that \emph{``the card fell short of the optimum aid in completing the task because it did not provide a full explanation of some of the material presented''}. Similarly, CP71, another compliance expert, stated that \emph{``the card was overall a helpful tool, but should provide more guidance on how to address complex aspects of the AI system''.} Participants recommended visual improvements to the card. For example, CP13, a developer, commented that \emph{``the legend at the bottom of each visualization should be moved closer to the top''.} Similarly, CP158, an ordinary individual, mentioned that \emph{``just filling in the box without explaining what it meant was not useful. It would have been more useful to have a rating with explanation for each category''}. We addressed comments on visual elements and provided the final versions of the card (Appendix~\ref{app:cards-after-updates}, Figures \ref{fig:updated-card-biometric-checkout}-\ref{fig:updated-card-license-plate-detector}). 

In our user study, we evaluated impact assessment cards for AI systems with a tangible presence in the physical world such as biometric checkout systems and license plate detectors. However, many algorithmic systems operate without a visible manifestation, for example, recommender systems or decision-support algorithms in public services and finance. To illustrate the adaptability of our card beyond physically situated AI, we created examples for two additional digital systems: a music recommender system \cite{recommenderSystem2013} (Appendix~\ref{app:card-digital-system}, Figure \ref{fig:card-music-recommender}) and a housing benefit allocation assistant \cite{tang2024ai} (Appendix~\ref{app:card-digital-system}, Figure \ref{fig:card-benefit-assistant}). These examples demonstrate how our card can incorporate different visual elements and be adapted for AI systems that operate in the background, often without end users being fully aware of their presence.
\section{Discussion}
\label{sec:discussion}
We begin by consolidating our findings on the use of impact assessment cards as tools for assessing AI risks, communicating AI benefits, and supporting AI governance (\S\ref{subsec:card_assessment}). Next, we explore opportunities to apply the cards in different contexts (\S\ref{subsec:context_appropriation}), and discuss their limitations (\S\ref{subsec:future_directions}).

\subsection{Cards as Tools for Assessing AI Risks, Communicating AI Benefits, and Supporting AI Governance}
\label{subsec:card_assessment}
The impact assessment card offers a new accessible medium for addressing the ethical and practical aspects of AI systems. Unlike detailed reports aimed at primarily technical audiences, our card can engage diverse stakeholders with a concise and visually appealing format. Next, we discuss three prospective applications of the card.
\smallskip

\noindent\textbf{Assessing AI Risks.} HCI and CSCW research has long emphasized the importance of tools that help stakeholders foresee potential failures and harms in technology design~\cite{hong2021planning, kozubaev2020expanding, constantinides2023method}. Our findings demonstrate that impact assessment cards enable stakeholders to identify and reflect on risks more effectively than full reports. Our participants engaged with the content, contextualizing risks in relation to AI uses and mitigations. By broadening access to risk discussions, impact assessment cards may foster informed decision-making and civic engagement in AI governance~\cite{buhmann2021towards, prabhakaran2022human, riskAtlas2024}.
\smallskip

\noindent\textbf{Communicating AI Benefits.} The public discourse on AI often emphasizes risks, overshadowing potential benefits~\cite{ojewale2024towards,novelli2023taking}. Our card aims at addressing this imbalance by presenting benefits prominently alongside risks, drawing inspiration from fields such as medicine and energy communication~\cite{InformationDesignLeaflet}. Our participants valued this balanced perspective, which encouraged deep reflections on the dual aspects of AI systems. This approach aligns with ethical principles of informed decision-making, ensuring that AI is seen as a tool with both opportunities and challenges~\cite{landgrebe2022machines}.
\smallskip

\noindent\textbf{Supporting AI Governance.} 
Existing governance tools (e.g., certification labels and audit frameworks) often target technically skilled users~\cite{scharowski2023certification, buccinca2023aha}. In contrast, our card synthesizes complex audit information into an accessible format suitable for a broader audience. This design decision is driven by the need for better alignment between technical experts and the broader public in AI governance~\cite{floridi2018ai4people}. Experts benefit from a concise tool for communicating governance decisions, while non-experts gain a practical resource that simplifies regulatory concepts and clarifies their rights. For example, engineers in AI companies could use the card for internal communication, while regulators might adopt it to support compliance with frameworks such as the EU AI Act~\cite{EUAIAct_2024}. Moreover, the cards can empower legal and civil society organizations by providing them with a user-friendly tool to engage in advocacy, oversight, and accountability efforts. By bridging the gap between technical and non-technical audiences, the cards advance inclusivity in AI governance.

\subsection{Cards Applied in Different Contexts}
\label{subsec:context_appropriation}
We view impact assessment cards as versatile tools adaptable to various domains and stakeholder needs. We outline design opportunities and potential applications of the cards across four contexts.
\smallskip

\noindent\textbf{Participatory Design and Stakeholder Engagement.} Participatory design methodologies (e.g., focus groups or co-design workshops~\cite{muller1993participatory}) can be used to further refine the cards, ensuring their relevance across diverse use cases. These activities can identify stakeholder-specific needs, ensuring that the resulting card addresses both direct and indirect impacts of AI systems~\cite{judgmentCall}. For example, in healthcare, impact assessment cards could include risk-benefit information tailored to AI-assisted diagnosis tools, highlighting concerns such as data privacy and patient safety, while showcasing benefits such as early detection of diseases. Similarly, in urban planning, the cards could map AI applications such as predictive traffic management, focusing on stakeholder groups such as residents, city planners, and policy makers.
\smallskip

\noindent\textbf{Regulatory and Compliance Applications.} Beyond summarizing risks and benefits, the cards could serve as templates for regulatory reporting, assisting organizations in mapping risks, mitigations, and benefits to regulatory requirements~\cite{AIActChecker}. By integrating data from datasheets and model cards~\cite{gebru2021datasheets, mitchell2019model}, impact assessment cards can help ensure transparency and accountability in AI governance. For example, an AI company developing a recruitment algorithm might customize the card to include categories such as bias mitigation strategies, compliance with anti-discrimination laws, and transparency measures. Including visual markers such as checkboxes or compliance level indicators (e.g., high, medium, low) could enhance their practicality for audits and self-assessments. Moreover, the cards could be integrated into certification processes, serving as an interface between technical audits and public-facing labels. For example, an AI certification body might use cards to communicate whether a system adheres to standards of transparency, fairness, or energy efficiency.
\mbox{}

\noindent\textbf{Educational and Advocacy Tools.}
The cards may also serve as tools for education and public advocacy. In academic settings, they can introduce students to the societal implications of AI through a structured way to explore ethical dilemmas~\cite{stavrakakis2021teaching}. By presenting complex topics in an accessible format, the cards help bridge the gap between technical knowledge and societal considerations, making them ideal for discussions on AI ethics, governance, and responsible innovation. Advocacy organizations could also employ the cards in public engagement campaigns to facilitate community discussions on AI-related issues such as data privacy, bias, and surveillance. The cards' utility and reach could be further expanded by integrating multimedia features such as QR codes linking to additional resources.
\smallskip

\noindent\textbf{Industry-Specific Appropriations.} 
Impact assessment cards can also be tailored to specific industries to address unique risks and benefits. In the financial sector, for example, they could evaluate AI-driven investment tools or fraud detection systems, focusing on transparency about decision-making criteria and biases. Similarly, in the energy and environment domain, the cards might highlight trade-offs in AI applications for renewable energy optimization, helping stakeholders balance gains in efficiency with risks related to system reliability and data accuracy.

Cards are also applicable across systems with varying levels of (physical) visibility. For example, physically situated AI systems (like our exemplary biometric checkout and license plate detector) can be seen as systems with material manifestation as they have a material presence in the physical world (at the point of checkout, or through cameras and CCTVs). On the contrary, AI systems without a material manifestation (e.g., music recommendation platforms \cite{recommenderSystem2013}, benefit allocation assistants \cite{tang2024ai}) operate entirely in digital environments where their presence is not tied to a physical location but rather integrated into software interfaces or cloud-based services. For systems with material manifestations, cards can provide clear information about data processing and privacy measures. For example, in a biometric checkout system that enables customers to make payments using facial recognition, cards could appear at key moments in the customer journey: during enrollment when users scan their face and link it to a payment method, or on receipts as a QR code to reinforce transparency after a transaction. Conversely, for systems without material manifestations such as recommender systems \cite{recommenderSystem2013}, cards can promote transparency about algorithms and biases. In platforms like Spotify or Netflix, cards could explain how recommendations are generated, including the use of data sources and personalization algorithms, and highlight any associated risks or biases (Appendix~\ref{app:card-digital-system}, Figure~\ref{fig:card-music-recommender}). These cards could be integrated into digital touchpoints such as during account setup alongside terms and conditions, or embedded in the platform's navigation bar under sections like ``About'' or ``Transparency''. By positioning the cards strategically, users can easily access and understand how their data is used, fostering trust and accountability.
\smallskip

\subsection{Limitations and Future Directions}
\label{subsec:future_directions}
Our study and the impact card have four main limitations that suggest directions for future research. First, its brevity may overlook the complexities of AI risks and benefits, requiring more research to adapt it for diverse real-world AI applications. 
Future designs could involve creating culturally varied card versions \cite{colorPsychology_2015}, or blending physical and digital forms with interactive elements for better risk and benefit understanding \cite{visWhatWorks2021}. Despite their potential, we believe that cards are not a replacement for detailed reports, particularly in contexts requiring comprehensive evidence to substantiate compliance claims. Participants recommended simplifying language, summarizing key points, and incorporating visual aids to make reports more accessible. Future work could explore how hybrid tools---combining cards and reports---might balance accessibility and depth, further enhancing stakeholder engagement. Second, although the card received higher usability ratings from all cohorts, design improvements could further enhance its usability. The card's score partly reflects the challenge of our endeavor: to create a user-friendly tool that effectively communicates the risks and benefits of AI in a way that is accessible to individuals without technical expertise. In the future, we plan to broaden our engagement to include a more diverse group of stakeholders such as organizational leaders. Third, our study's sample may not completely represent all AI stakeholders like developers, compliance experts, and ordinary individuals due to limited controls over participants' roles, AI use frequency, and location. While we recruited a sample of Prolific participants matching the US population, the findings and discussions should be interpreted with some limitations. For example, our study does not account for the recently updated standards introducing a combined race and ethnicity question where groups such as Hispanic or Latino are considered a co-equal category alongside the ethnicity categories we used \cite{newEthnicityClassifications_2024}. We encourage future researchers to expand ethnicity screening when recruiting participants to improve representation.

Finally, AI students, crucial for future AI development \cite{johnson2023classroomStudy}, were not part of our cohorts. Inspired by research on the AI Incident Database's educational impact \cite{feffer2023ai}, we aim to integrate impact cards with incident reports in future studies to assess AI students' understanding of risks and benefits.
\section{Conclusion}
\label{sec:conclusion}
Through an iterative design process, we developed and evaluated an impact assessment card for communicating the risks and benefits of AI uses. The card summarizes detailed AI reports, presenting complex information in a clear and accessible way for both experts and laypeople. We evaluated our card's effectiveness in an online study with 235 participants across developers, compliance experts, and ordinary individuals. We found that the card's effectiveness extended beyond ordinary individuals, offering advantages even to those already well-versed in AI impact assessments. Moving forward, our work suggests a promising direction for further refining impact assessment cards, aiming to democratize understanding and participation in AI impact assessment~\cite{erman2024democratization}.

\begin{acks}
This work was done at Nokia Bell Labs. MC was supported by Nokia Bell Labs and the European Union's Horizon 2020 Research and Innovation Programme (grant agreement No. 739578).
\end{acks}

\newpage

\bibliographystyle{ACM-Reference-Format}
\bibliography{main}

\appendix
\clearpage
\section{Appendix}
\subsection{Design Patterns for Communicating Risks and Benefits of AI Uses}
\label{app:design-patterns}

\begin{figure*}[h!]
  \flushleft
  \includegraphics[width=0.79\textwidth]{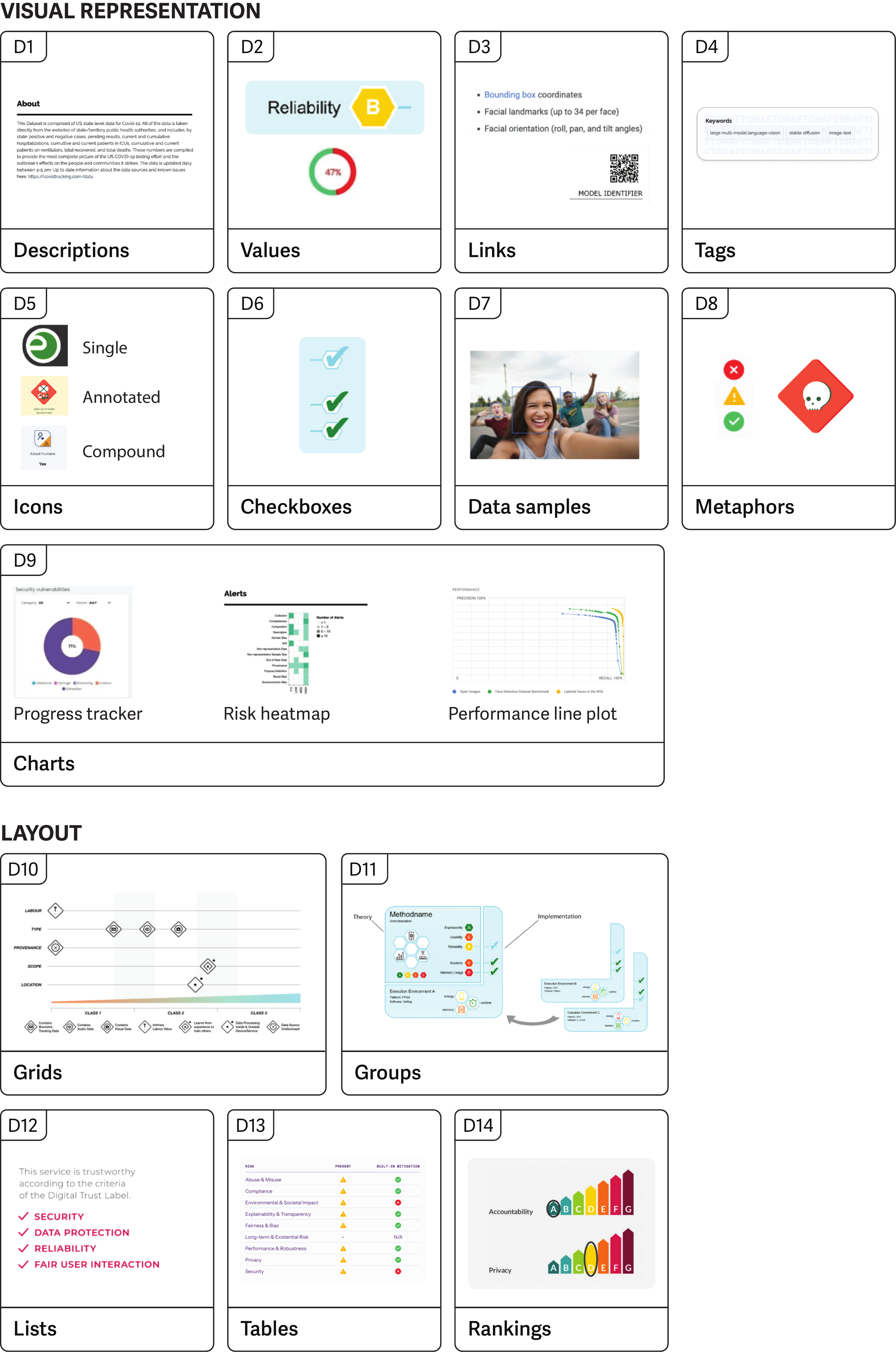}
  \caption{Fourteen design patterns for \textbf{visual representation} (D1-D9) and \textbf{layout} (D10-D14) to communicate the risks and benefits of AI technologies, derived from the literature review~\cite{cihon-ai-certification-2021, stuurman2021, ojewale2024towards, holland-nutritionLabel-2018, lindley-aiLegibility-2020, morik-careLabel-2022, gorton2021determines, jones2019front, stadelmann2018different, data_hazards}. Also available at: \url{https://social-dynamics.net/ai-risks/impact-card/}.}
  \label{fig:design-patterns}
\end{figure*}

\clearpage
\subsection{Impact Assessment Cards Used in a Large-Scale Online Study}
\label{app:cards-user-study}

\begin{figure*}[h!]
  \centering
  \includegraphics[width=0.79\textwidth]{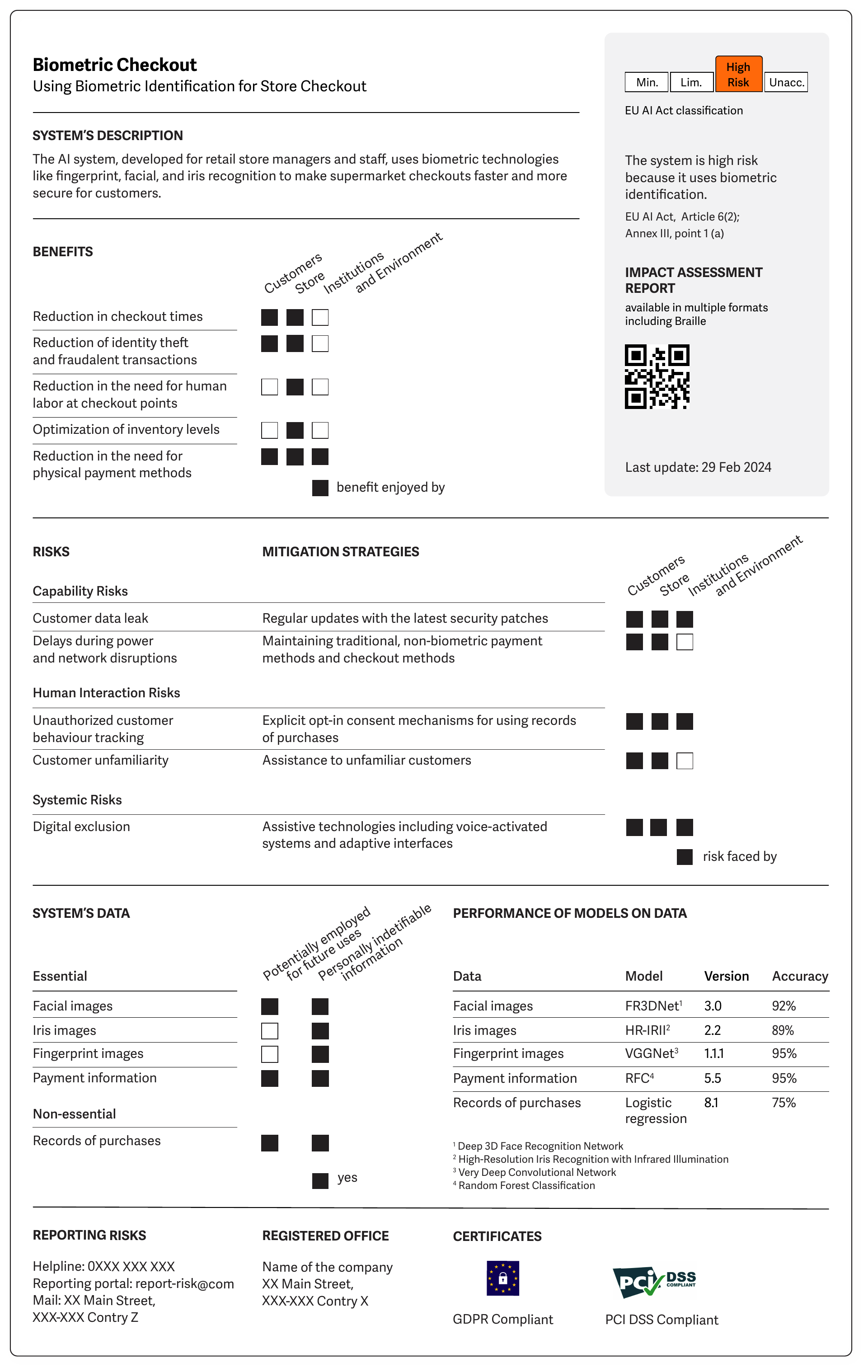}
  \caption{Impact assessment card for a store checkout system using biometric identification, used during the large-scale online study. Also available at: \url{https://social-dynamics.net/ai-risks/impact-card/}.}
  \label{fig:card-biometric-checkout}
\end{figure*}

\begin{figure*}[h!]
  \centering
  \includegraphics[width=0.79\textwidth]{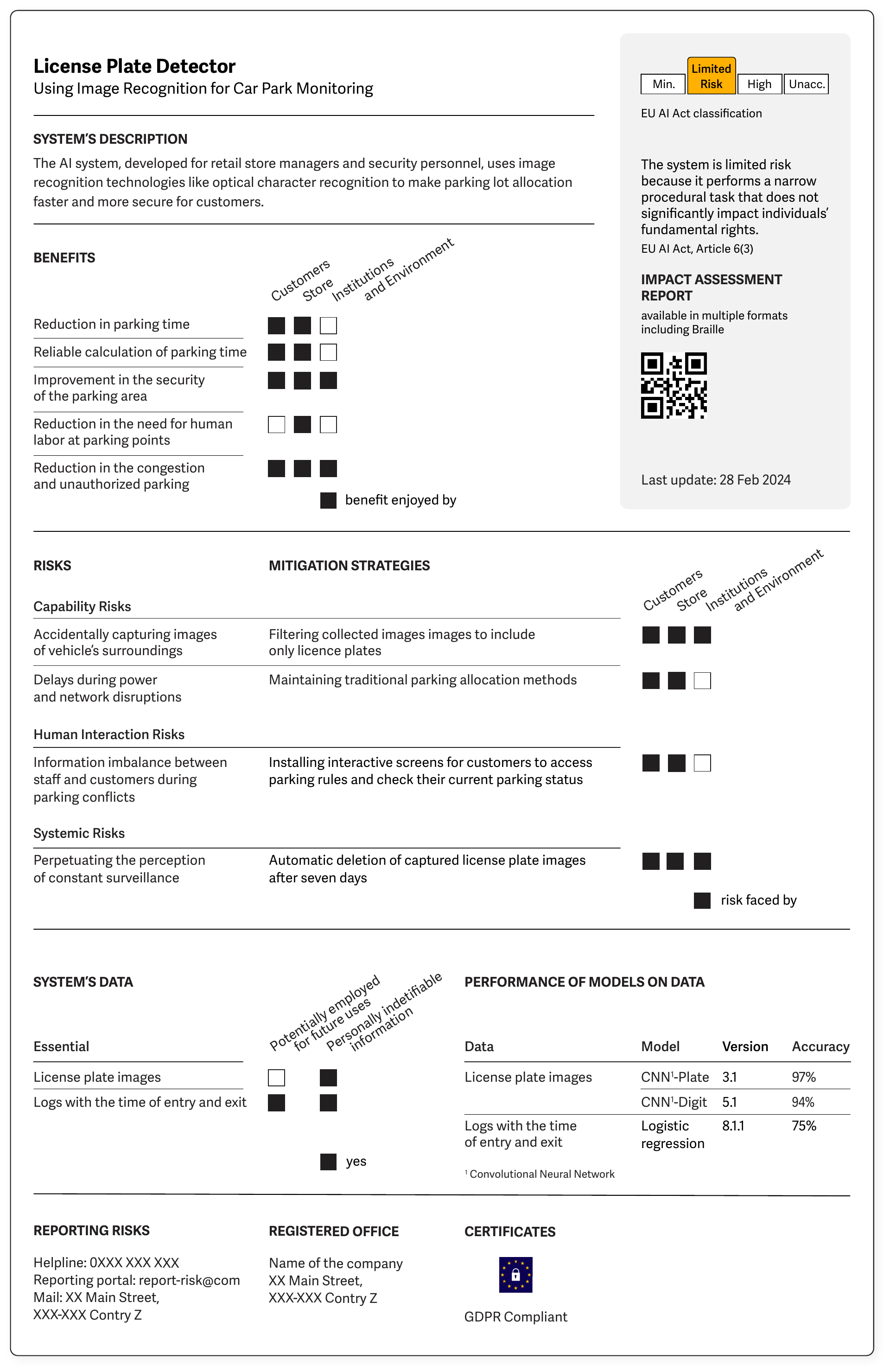}
  \caption{Impact assessment card for a car park monitoring system using image recognition, used during the large-scale online study. Also available at: \url{https://social-dynamics.net/ai-risks/impact-card/}.}
  \label{fig:card-license-plate-detector}
\end{figure*}

\clearpage
\subsection{Impact Assessment Reports Used in a Large-Scale Online Study}
\label{app:reports-user-study}

\begin{figure*}[h!]
  \centering
  \includegraphics[width=\textwidth]{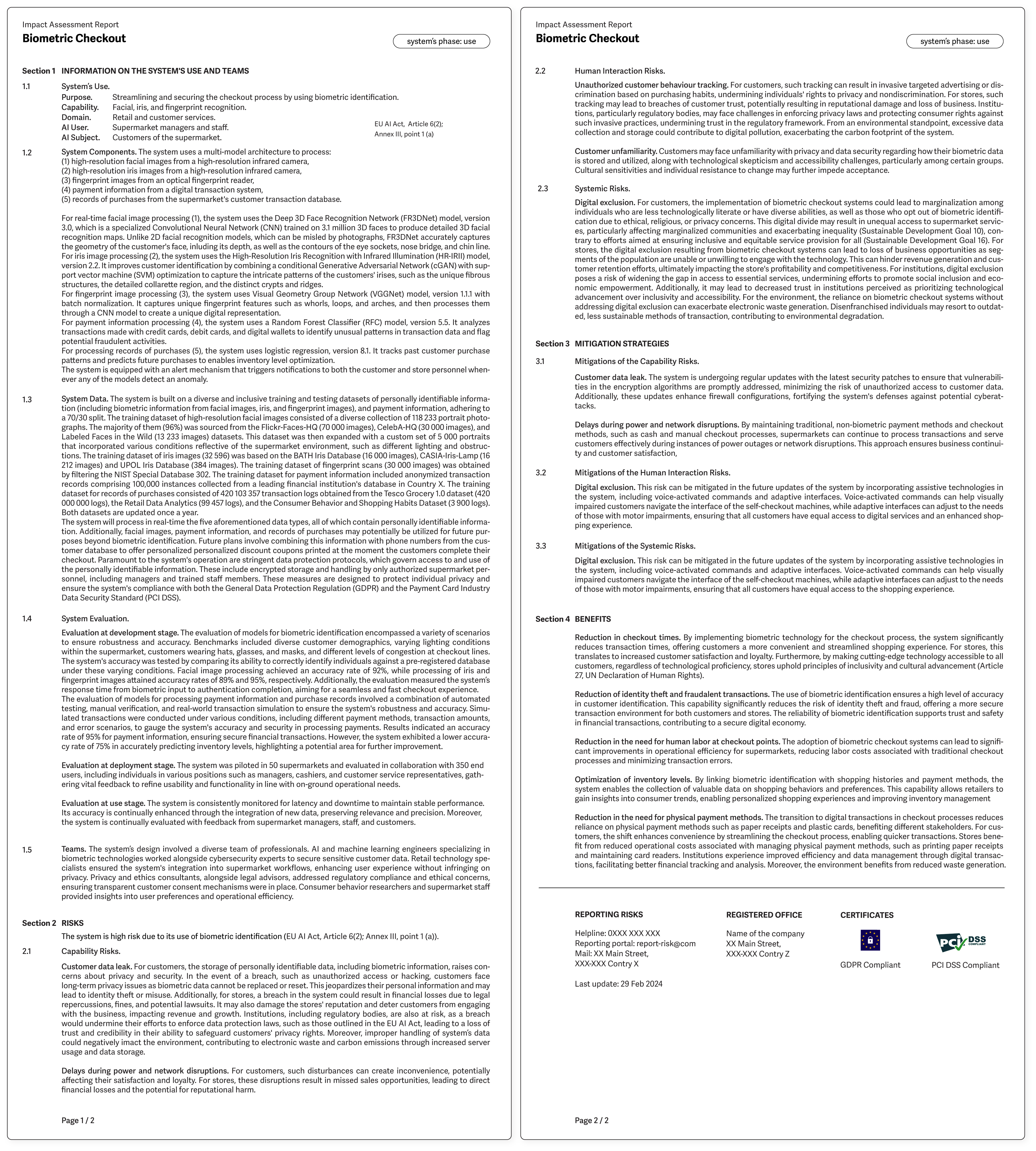}
  \caption{Impact assessment report for a store checkout system using biometric identification, used during the large-scale online study. Also available at:\url{https://social-dynamics.net/ai-risks/impact-card/}.}
  \label{fig:report-biometric-checkout}
\end{figure*}

\begin{figure*}[h!]
  \centering
  \includegraphics[width=\textwidth]{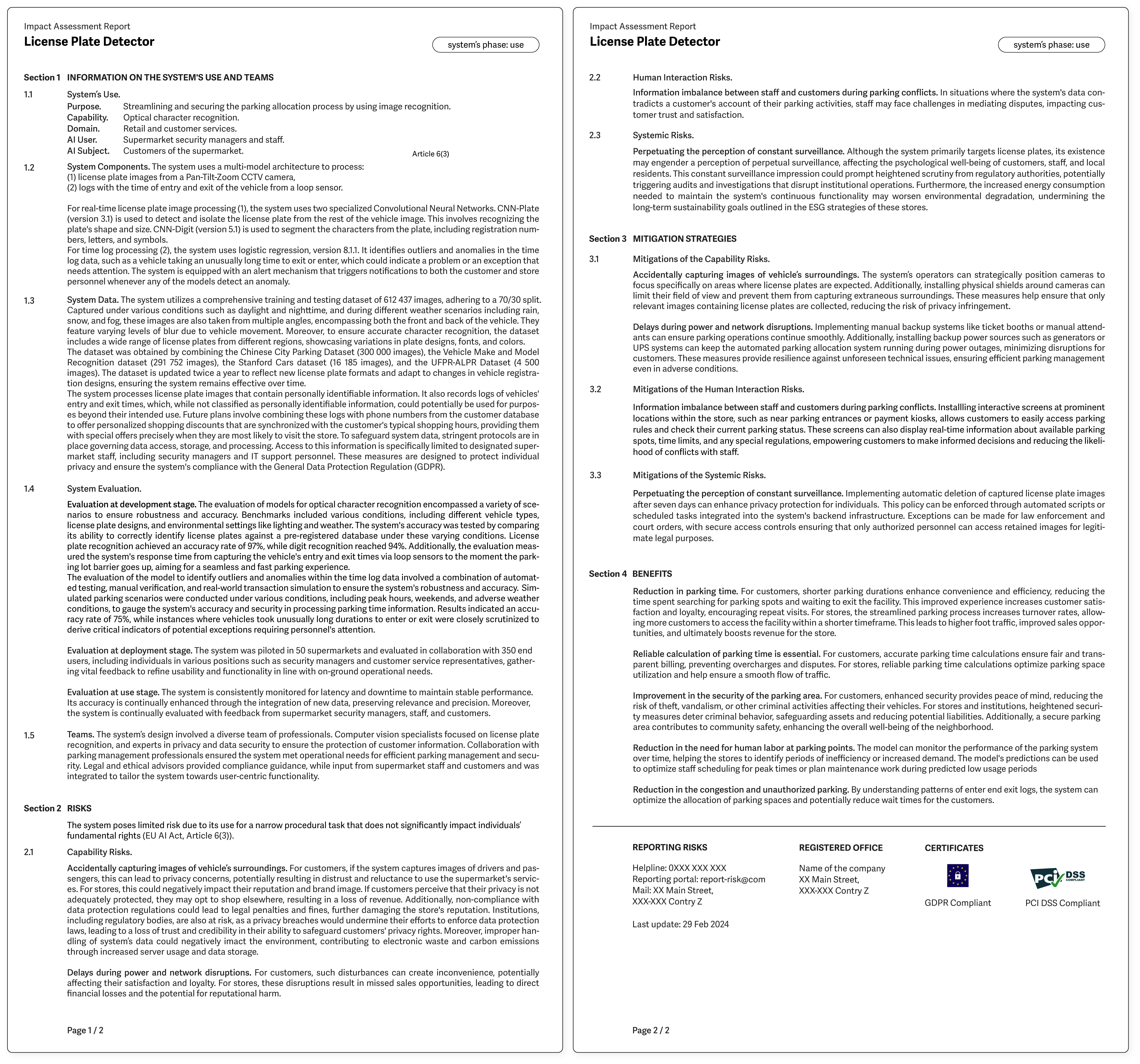}
  \caption{Impact assessment report for a car park monitoring system using image recognition, used during the large-scale online study. Also available at: \url{https://social-dynamics.net/ai-risks/impact-card/}.}
  \label{fig:report-license-plate-detector}
\end{figure*}

\clearpage
\subsection{Rubric for Evaluating Email Quality}
\label{app:evaluation-rubric}

The quality of emails was assessed based on five key criteria. Each email was evaluated for its ability to: address the real-world use of the system (\emph{context}), provide a clear call to action for or against implementation (\emph{recommendation}), identify and discuss risks associated with the system (\emph{risks}), mention actionable strategies to mitigate these risks (\emph{mitigations}), and present information in a clear and coherent manner (\emph{content clarity}). Lower-rated emails often failed to directly engage with the system described and were vague. Higher-rated emails demonstrated a nuanced understanding of the system, offered balanced arguments covering risks and benefits, and included solutions for addressing identified risks.

\vspace{0.35cm}
\noindent\textbf{Detailed Criteria for Email Quality Ratings}

\vspace{0.6cm}
\tikzstyle{background rectangle}=[thick, draw=black, rounded corners]
\begin{tikzpicture}[show background rectangle]
\node[align=justify, text width=35.5em, inner sep=1em]{
    \noindent\textbf{Context:} No mention of the system's real-world use. \\
    \noindent\textbf{Recommendation:} Lacks a decisive recommendation. \\
    \noindent\textbf{Risks:} Fails to mention any risks. \\
    \noindent\textbf{Mitigations:} Does not include any mitigation strategies or references to actions for reducing risks. \\
    \noindent\textbf{Content clarity:} Content is highly vague or incomprehensible. Structure or grammar issues significantly hinder readability.
};
\node[xshift=0.5ex, yshift=1ex, overlay, fill=black, text=white, draw=black, rounded corners, right=2.02cm, below=-0.3cm, inner xsep=0.55em, inner ysep=0.32em] at (current bounding box.north west) {
\textit{Rating 1: Poor quality}
};
\end{tikzpicture}

\vspace{0.65cm}
\tikzstyle{background rectangle}=[thick, draw=black, rounded corners]
\begin{tikzpicture}[show background rectangle]
\node[align=justify, text width=35.5em, inner sep=1em]{

    \noindent\textbf{Context:} Briefly mentions the system's real-world use but lacks elaboration. \\
    \noindent\textbf{Recommendation:} The recommendation is unclear or weak. \\
    \noindent\textbf{Risks:} Includes at least one risk. Risks are mentioned but lack relevance to the system's real-world use. \\
    \noindent\textbf{Mitigations:} Includes at least one mitigation strategy. Mitigation strategies are mentioned but are not clearly tied to the specific risks of the system's real-world use. \\
    \noindent\textbf{Content clarity:} Content is somewhat vague or challenging to follow. Structure lacks focus and clarity.
};
\node[xshift=0.5ex, yshift=1ex, overlay, fill=black, text=white, draw=black, rounded corners, right=1.98cm, below=-0.3cm, inner xsep=0.55em, inner ysep=0.32em] at (current bounding box.north west) {
\textit{Rating 2: Fair quality}
};
\end{tikzpicture}

\vspace{0.65cm}
\tikzstyle{background rectangle}=[thick, draw=black, rounded corners]
\begin{tikzpicture}[show background rectangle]
\node[align=justify, text width=35.5em, inner sep=1em]{

    \noindent\textbf{Context:} Explains the system's real-world use in at least one sentence, demonstrating a basic understanding of the system. \\
    \noindent\textbf{Recommendation:} The recommendation is clear (recommend or reject) but could be more compelling. \\
    \noindent\textbf{Risks:} Includes at least two risks. Risks are moderately connected to the system. \\
    \noindent\textbf{Mitigations:} Includes at least two mitigation strategies tied to specific risks of the system's real-world use. \\
    \noindent\textbf{Content clarity:} Content is well-written with minimal vagueness. Logical structure supports the argument, though it may lack sophistication.
};
\node[xshift=0.5ex, yshift=1ex, overlay, fill=black, text=white, draw=black, rounded corners, right=2.07cm, below=-0.3cm, inner xsep=0.55em, inner ysep=0.32em] at (current bounding box.north west) {
\textit{Rating 3: Good quality}
};
\end{tikzpicture}

\vspace{0.65cm}
\vspace*{0.15cm}
\tikzstyle{background rectangle}=[thick, draw=black, rounded corners]
\begin{tikzpicture}[show background rectangle]
\node[align=justify, text width=35.5em, inner sep=1em]{

    \noindent\textbf{Context:} Explains the system's real-world use in detail, demonstrating a clear grasp of its operational implications and relevance to its users and subjects. \\
    \noindent\textbf{Recommendation:} The recommendation is clear (recommend or reject) and decisive. \\
    \noindent\textbf{Risks:} Identifies and discusses multiple risks (>2). Risks are clearly connected to the system. Attempts to prioritize key risks. \\
    \noindent\textbf{Mitigations:} Includes more than two mitigation strategies tied to specific risks of the system's real-world use. The included mitigations are actionable. \\
    \noindent\textbf{Content clarity:} Content is very well-written and logically structured, making the information easy to follow. Arguments are cohesive and well-supported.
};
\node[xshift=0.5ex, yshift=1ex, overlay, fill=black, text=white, draw=black, rounded corners, right=2.4cm, below=-0.3cm, inner xsep=0.55em, inner ysep=0.32em] at (current bounding box.north west) {
\textit{Rating 4: Very good quality}
};
\end{tikzpicture}

\vspace{0.65cm}
\tikzstyle{background rectangle}=[thick, draw=black, rounded corners]
\begin{tikzpicture}[show background rectangle]
\node[align=justify, text width=35.5em, inner sep=1em]{
    
    \noindent\textbf{Context:} Demonstrates a nuanced understanding of the system's real-world use, with concrete examples and scenarios related to its users and subjects. \\
    \noindent\textbf{Recommendation:} The recommendation is clear (recommend or reject) and decisive. It balances pros and cons with depth and foresight. \\
    \noindent\textbf{Risks:} Identifies and thoroughly discusses all key risks, including subtle or rare risks, or identifies new risks expanding the scope of the treatment. Effectively prioritizes risks with clear justification. \\
    \noindent\textbf{Mitigations:} Includes more than two mitigation strategies tied to specific risks of the system's real-world use. The included mitigations are actionable, specific, and technically detailed. \\
    \noindent\textbf{Content clarity:} Exceptionally clear, precise, and insightful writing style. Engages the reader and delivers a compelling, logical argument.
};
\node[xshift=0.5ex, yshift=1ex, overlay, fill=black, text=white, draw=black, rounded corners, right=2.35cm, below=-0.3cm, , inner xsep=0.55em, inner ysep=0.32em] at (current bounding box.north west) {
\textit{Rating 5: Excellent quality}
};
\end{tikzpicture}

\clearpage
\noindent\textbf{Examples of Emails and Their Quality Ratings}

\vspace{0.65cm}
\tikzstyle{background rectangle}=[thick, draw=black, rounded corners]
\begin{tikzpicture}[show background rectangle]
\node[align=justify, text width=35.5em, inner sep=1em]{

    \noindent\textbf{Email text:} I hope this email finds you and your team doing well. I recently stumbled upon and read your document implementing the new AI system in the store. This will definitely help us improve our understanding and familiarity with artificial intelligence. \\
    \noindent\textbf{Justification for rating:} The email lacks specific references to the AI system's use, does not state a recommendation or rejection, and fails to provide arguments related to the system's use, risks, or mitigations.
};
\node[xshift=0.5ex, yshift=1ex, overlay, fill=black, text=white, draw=black, rounded corners, right=2.02cm, below=-0.3cm, inner xsep=0.55em, inner ysep=0.32em] at (current bounding box.north west) {
\textit{Rating 1: Poor quality}
};
\end{tikzpicture}

\vspace{0.65cm}
\tikzstyle{background rectangle}=[thick, draw=black, rounded corners]
\begin{tikzpicture}[show background rectangle]
\node[align=justify, text width=35.5em, inner sep=1em]{

    \noindent\textbf{Email text:} Dear team, it has come to my attention that there are significant risk associated with the use of the biometric checkout system. Considering these risks and the potential negative impact on both customers and the environment, I kindly request that we consider the re-implementation of the system. \\
    \noindent\textbf{Justification for rating:} The email references the specific AI system's use, but the recommendation is unclear. It mentions general risks and includes re-development as one mitigation strategy.
};
\node[xshift=0.5ex, yshift=1ex, overlay, fill=black, text=white, draw=black, rounded corners, right=1.98cm, below=-0.3cm, inner xsep=0.55em, inner ysep=0.32em] at (current bounding box.north west) {
\textit{Rating 2: Fair quality}
};
\end{tikzpicture}

\vspace{0.65cm}
\tikzstyle{background rectangle}=[thick, draw=black, rounded corners]
\begin{tikzpicture}[show background rectangle]
\node[align=justify, text width=35.5em, inner sep=1em]{

    \noindent\textbf{Email text:} Hello, I'm writing this email to you to recommend implementing the system of biometric checkout. Although, a good and safe option would be to make an opt-in system and inform the customers. That way people can't complain. I understand some people will say this is too much surveillance and will invade our privacy. Overall, it will reduce identity theft and fraudulent transactions. Let's say, you lose your card, someone else grabs it and tries to buy something at a store, with biometric checkout, this person will be caught. \\
    \noindent\textbf{Justification for rating:} The email references the specific use of the AI system, and its recommendation is clear. It identifies two primary risks associated with the system: the potential for surveillance and privacy invasion. To address these risks, the email outlines two specific mitigation strategies: implementing an opt-in system to ensure user consent and proactively informing customers about how their data will be used. Additionally, the email provides a balanced perspective by including arguments that highlight both the potential benefits and drawbacks of the system.
};
\node[xshift=0.5ex, yshift=1ex, overlay, fill=black, text=white, draw=black, rounded corners, right=2.07cm, below=-0.3cm, inner xsep=0.55em, inner ysep=0.32em] at (current bounding box.north west) {
\textit{Rating 3: Good quality}
};
\end{tikzpicture}

\vspace{1.65cm}
\vspace*{0.15cm}
\tikzstyle{background rectangle}=[thick, draw=black, rounded corners]
\begin{tikzpicture}[show background rectangle]
\node[align=justify, text width=35.5em, inner sep=1em]{
    \noindent\textbf{Email text:} Dear Ethics Committee, I am writing this email to advise against implementing this license plate detector for use in the parking lot. I think the potential risks with this AI system outweighs the potential benefits. I believe that accidentally capturing images of vehicle's surroundings will lead to conflict between staff and customers, as well as raise privacy concerns among customers. The mitigation strategies included, such as installing physical shields and shortening data storage time, will increase cost for a solution that doesn't help much. The benefits of this system are more for the store's benefit rather than the customers', which could potentially lead to losing customers. \\
    \noindent\textbf{Justification for rating:} The email provides a detailed explanation of the AI system's real-world use, showcasing a strong understanding of its operational implications and relevance to both customers and the store. The recommendation is clear and decisive, addressing four key risks: conflicts between staff and customers, privacy concerns among customers, potential loss of customers, and increased costs for the store. It offers more than two targeted mitigation strategies tied to these risks, presented in a well-written and logically structured manner.
};
\node[xshift=0.5ex, yshift=1ex, overlay, fill=black, text=white, draw=black, rounded corners, right=2.4cm, below=-0.3cm, inner xsep=0.55em, inner ysep=0.32em] at (current bounding box.north west) {
\textit{Rating 4: Very good quality}
};
\end{tikzpicture}

\vspace{0.65cm}
\tikzstyle{background rectangle}=[thick, draw=black, rounded corners]
\begin{tikzpicture}[show background rectangle]
\node[align=justify, text width=35.5em, inner sep=1em]{

    \noindent\textbf{Email text:} Dear Ethics Committee, I have reviewed the information presented and I *advise against* adopting this system. The following risks are described: (1) Risk of customer data leak - it is unacceptable to collect sensitive data such as biometrics and put it at risk of being stolen or leaked. ``Latest security patches'' is not sufficient as a strategy to safeguard high-value data, especially if your organization is specifically targeted for data theft. (2) Unauthorized customer tracking - there is no guarantee that a hacker, rogue employee, or rogue vendor cannot access the data and use it for their own purposes, even when safeguards exist. (3) Customer unfamiliarity - What if customers become familiar through the materials ans and don't want to use it? They will stop using the store, or they will complain on social media or to government regulators. The benefit does not outweigh the risk. In short, you must judge the system by what happens when it fails, not when everything goes right. In this case the legal and ethical risk is too high when the system fails. \\
    \noindent\textbf{Justification for rating:} The email provides a clear and well-justified recommendation. It explains the AI system's real-world use in detail, demonstrating concrete scenarios beyond those explicitly mentioned in the treatment, related to affected individuals. The recommendation is clear and decisive, prioritizing three key risks and including three targeted mitigation strategies directly tied to these risks.
};
\node[xshift=0.5ex, yshift=1ex, overlay, fill=black, text=white, draw=black, rounded corners, right=2.35cm, below=-0.3cm, , inner xsep=0.55em, inner ysep=0.32em] at (current bounding box.north west) {
\textit{Rating 5: Excellent quality}
};
\end{tikzpicture}

\clearpage
\subsection{Results of Regression Analyses for Predicting Task Completion Time, Usability Ratings, and Preference for Cards vs. Reports.}
\label{app:other-regressions}

\subsubsection{Predicting Task Completion Time}
\noindent\textbf{}

\begin{table*}[h!]
\flushleft
\small
\caption{\textbf{Factors influencing completion time include treatment and participant's cohort.} Using the report significantly increases completion time compared to the card, while legal experts take longer to complete the task than ordinary individuals. We conducted an ordinary least squares regression analysis with completion time as the dependent variable. The coefficients represent the changes in completion time (in seconds) relative to the reference category, with statistical significance indicated by: * for $p<0.05$, and ** for $p<0.01$. Non-significant factors (p > 0.05) are also reported for completeness.}
\label{tab:regression-completion}
\scalebox{0.87}{
\begin{tabular}{p{3.1cm}|p{6.4cm}|p{2cm}|p{2.95cm}}
\toprule
\textbf{Factor} & \textbf{Comparison (\emph{vs.} Reference Category)} & \textbf{Coefficient} & \textbf{$p$-value} \\
\midrule
Intercept &  & 302.870 & 0.000*** \\
\hline
\multicolumn{4}{l}{\textbf{Type of task}}  \\
\hline
System & Plate Detector \emph{vs.} Checkout & -3.409 & 0.908 \\
\hline
\multicolumn{4}{l}{\textbf{Participant's cohort}}  \\
\hline
Cohort & Developers \emph{vs.} Ordinary individuals & -50.423 & 0.200 \\
\cellcolor{highlight}\textbf{Cohort} & \cellcolor{highlight}\textbf{Legal Experts \emph{vs.} Ordinary individuals} & \cellcolor{highlight}\textbf{105.228} & \cellcolor{highlight}\textbf{0.013*} \\
\hline
\multicolumn{4}{l}{\textbf{Expertise levels}}  \\
\hline
Task Expertise & High \emph{vs.} Low & 13.841 & 0.474 \\
Technological Expertise & High \emph{vs.} Low & -26.685 & 0.227 \\
AI Expertise & High \emph{vs.} Low & 34.298 & 0.119 \\
\hline
\multicolumn{4}{l}{\textbf{Treatment}}  \\
\hline
\cellcolor{highlight}\textbf{Treatment type}  & \cellcolor{highlight}\textbf{Report \emph{vs.} Card} & \cellcolor{highlight}\textbf{102.617} & \cellcolor{highlight}\textbf{0.001**} \\
\bottomrule
\end{tabular}}
\end{table*}

\subsubsection{Predicting Usability Ratings}
\noindent\textbf{}

\begin{table*}[h!]
\flushleft
\small
\caption{\textbf{Factors influencing usability ratings include treatment and participant's cohort.} Participants across all cohorts find the report less usable than the card, and ordinary individuals give lower usability ratings compared to developers and legal experts. We conducted an ordinary least squares regression analysis with usability as the dependent variable. The coefficients represent the changes in usability scores relative to the reference category, with statistical significance indicated by: * for $p<0.05$, ** for $p<0.01$, and *** for $p<0.001$. Non-significant factors (p > 0.05) are also reported for completeness.}

\label{tab:regression-usability}
\scalebox{0.87}{
\begin{tabular}{p{3.1cm}|p{6.4cm}|p{2cm}|p{2.95cm}}
\toprule
\textbf{Factor} & \textbf{Comparison (\emph{vs.} Reference Category)} & \textbf{Coefficient} & \textbf{$p$-value} \\
\midrule
Intercept &  & 50.124 & 0.000 \\
\hline
\multicolumn{4}{l}{\textbf{Type of task}} \\
\hline
System & Plate Detector \emph{vs.} Checkout & 3.076 & 0.083 \\
\hline
\multicolumn{4}{l}{\textbf{Participant's cohort}}  \\
\hline
\cellcolor{highlight}\textbf{Cohort} & \cellcolor{highlight}\textbf{Developers \emph{vs.} Ordinary individuals}  & \cellcolor{highlight}\textbf{4.769} & \cellcolor{highlight}\textbf{0.043*} \\
\cellcolor{highlight}\textbf{Cohort} & \cellcolor{highlight}\textbf{Legal Experts \emph{vs.} Ordinary individuals} & \cellcolor{highlight}\textbf{6.199} & \cellcolor{highlight}\textbf{0.004**} \\
\hline
\multicolumn{4}{l}{\textbf{Expertise levels}}  \\
\hline
Task Expertise & High \emph{vs.} Low & 1.425 & 0.218 \\
Technological Expertise & High \emph{vs.} Low & 2.492 & 0.060 \\
AI Expertise & High \emph{vs.} Low & -0.776 & 0.555 \\
\hline
\multicolumn{4}{l}{\textbf{Treatment}}  \\
\hline
\cellcolor{highlight}\textbf{Treatment type}  & \cellcolor{highlight}\textbf{Report \emph{vs.} Card} & \cellcolor{highlight}\textbf{-11.750} & \cellcolor{highlight}\textbf{0.000***} \\
\bottomrule
\end{tabular}}
\end{table*}

\clearpage
\subsubsection{Predicting Preference for Cards vs. Reports}
\noindent\textbf{}

\begin{table*}[h!]
\flushleft
\small
\caption{\textbf{Factors influencing preference for cards include recommendation type, participant's cohort, and AI expertise.} ``Reject'' and ``Unclear'' recommendations, belonging to the legal experts cohort, and greater AI expertise all reduce the likelihood of preferring cards. We conducted a binomial logistic regression analysis with preference for cards (\emph{vs.} reports) as the dependent variable. The coefficients represent the log-odds of preferring a card relative to a report for each factor, compared to its reference category. Statistical significance is indicated by: * for $p<0.05$, and ** for $p<0.01$. Non-significant factors (p > 0.05) are also reported for completeness.}

\clearpage
\label{tab:regression-preference}
\scalebox{0.87}{
\begin{tabular}{p{3.1cm}|p{6.4cm}|p{2cm}|p{2.95cm}}
\toprule
\textbf{Factor} & \textbf{Comparison (\emph{vs.} Reference Category)} & \textbf{Coefficient} & \textbf{$p$-value} \\
\midrule
Intercept &  & 1.635 & 0.001 \\ \hline
\multicolumn{4}{l}{\textbf{Type of task}}  \\ \hline
\cellcolor{highlight}\textbf{Recommendation} & \cellcolor{highlight}\textbf{Reject \emph{vs.} Recommend} & \cellcolor{highlight}\textbf{-0.542} & \cellcolor{highlight}\textbf{0.018*} \\
\cellcolor{highlight}\textbf{Recommendation} & \cellcolor{highlight}\textbf{Unclear \emph{vs.} Recommend} & \cellcolor{highlight}\textbf{-0.936} & \cellcolor{highlight}\textbf{0.002**} \\
System & Plate Detector \emph{vs.} Checkout & -0.012 & 0.953 \\ \hline
\multicolumn{4}{l}{\textbf{Participant's cohort}}  \\ \hline
Cohort & Developers \emph{vs.} Ordinary individuals & -0.440 & 0.105 \\
\cellcolor{highlight}\textbf{Cohort} & \cellcolor{highlight}\textbf{Legal Experts \emph{vs.} Ordinary individuals} & \cellcolor{highlight}\textbf{-0.707} & \cellcolor{highlight}\textbf{0.005**} \\ \hline
\multicolumn{4}{l}{\textbf{Expertise levels}}  \\ \hline
Task Expertise & High \emph{vs.} Low & 0.151 & 0.258 \\
Technological Expertise & High \emph{vs.} Low & 0.178 & 0.240 \\
\cellcolor{highlight}\textbf{AI Expertise} & \cellcolor{highlight}\textbf{Low \emph{vs.} High} & \cellcolor{highlight}\textbf{-0.477} & \cellcolor{highlight}\textbf{0.002**} \\ \hline
\multicolumn{4}{l}{\textbf{Treatment}}  \\ \hline
Treatment type & Report \emph{vs.} Card & 0.132 & 0.523 \\
\bottomrule
\end{tabular}}
\end{table*}

\clearpage
\subsection{Final Impact Assessment Cards for Physically Situated AI Systems}
\label{app:cards-after-updates}

\begin{figure*}[h!]
  \centering
  \includegraphics[width=0.62\textwidth]{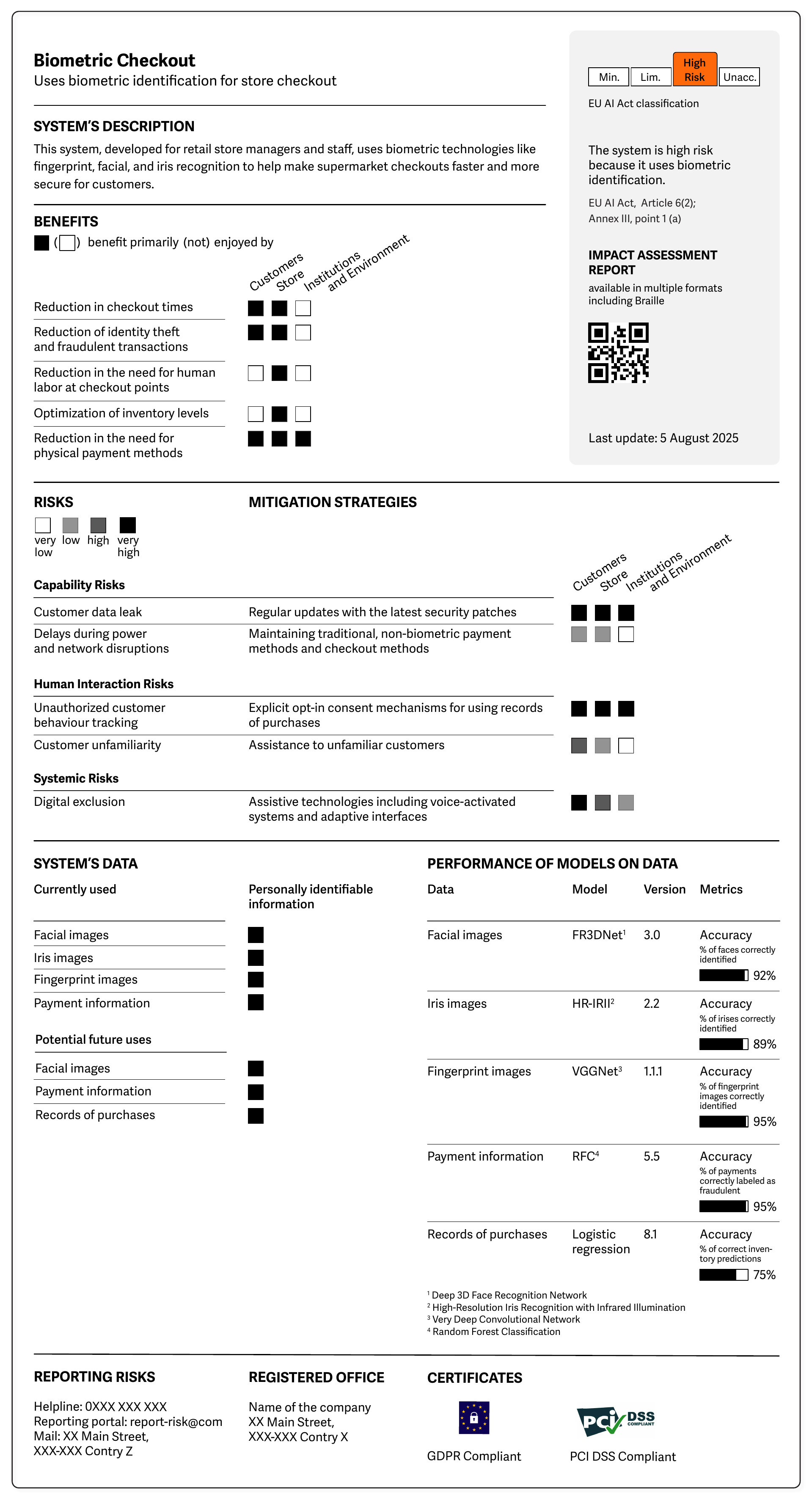}
  \caption{Final impact assessment card for a store checkout system using biometric identification, including the risk severity ratings. Also available at: \url{https://social-dynamics.net/ai-risks/impact-card/}.}
  \label{fig:updated-card-biometric-checkout}
\end{figure*}

\begin{figure*}[h!]
  \centering
  \includegraphics[width=0.62\textwidth]{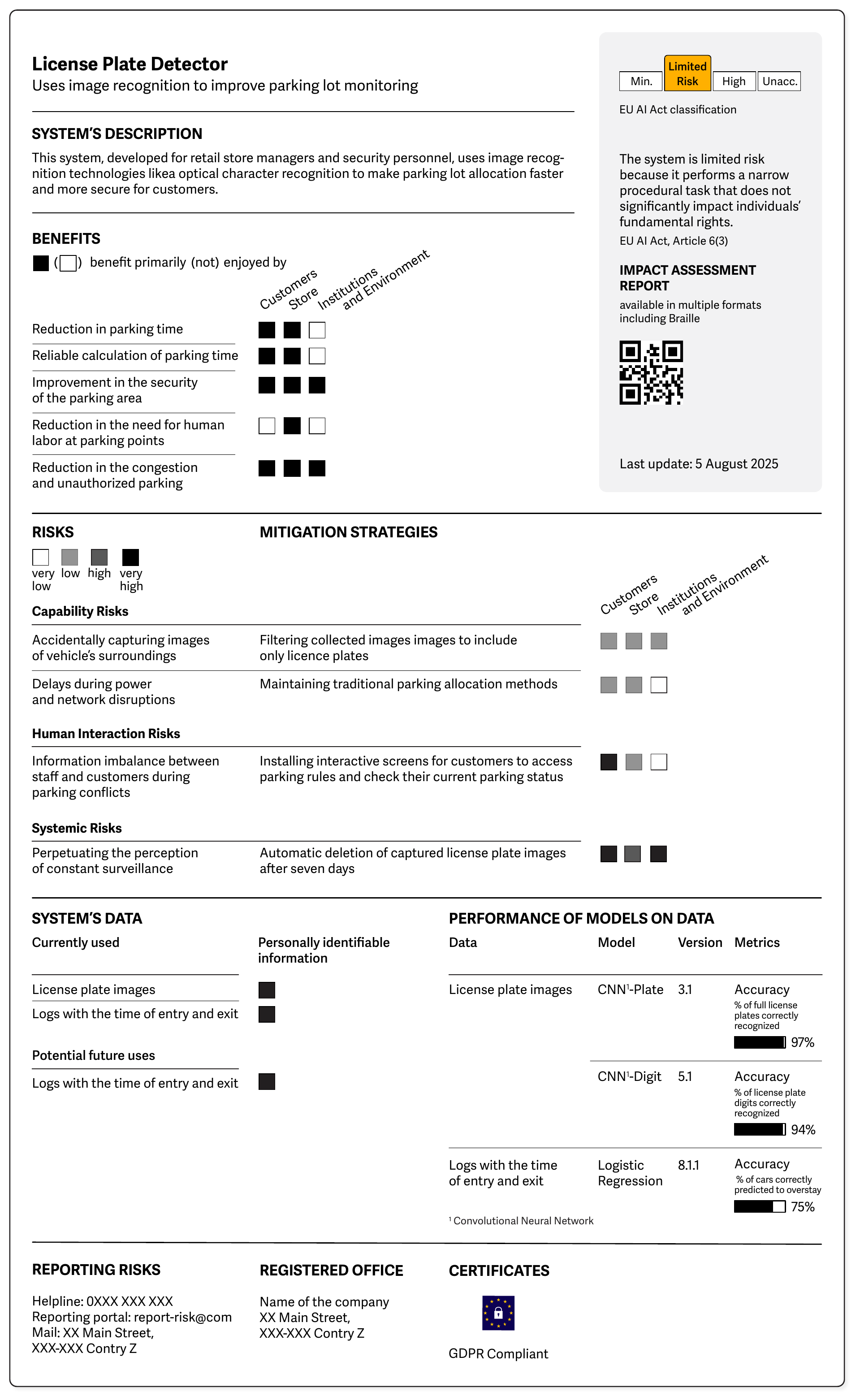}
  \caption{Final impact assessment card for a car park monitoring system using image recognition, including the risk severity ratings. Also available at: \url{https://social-dynamics.net/ai-risks/impact-card/}.}
  \label{fig:updated-card-license-plate-detector}
\end{figure*}

\clearpage
\subsection{Final Impact Assessment Cards for Digital AI Systems}
\label{app:card-digital-system}

\begin{figure*}[h!]
  \centering
  \includegraphics[width=0.62\textwidth]{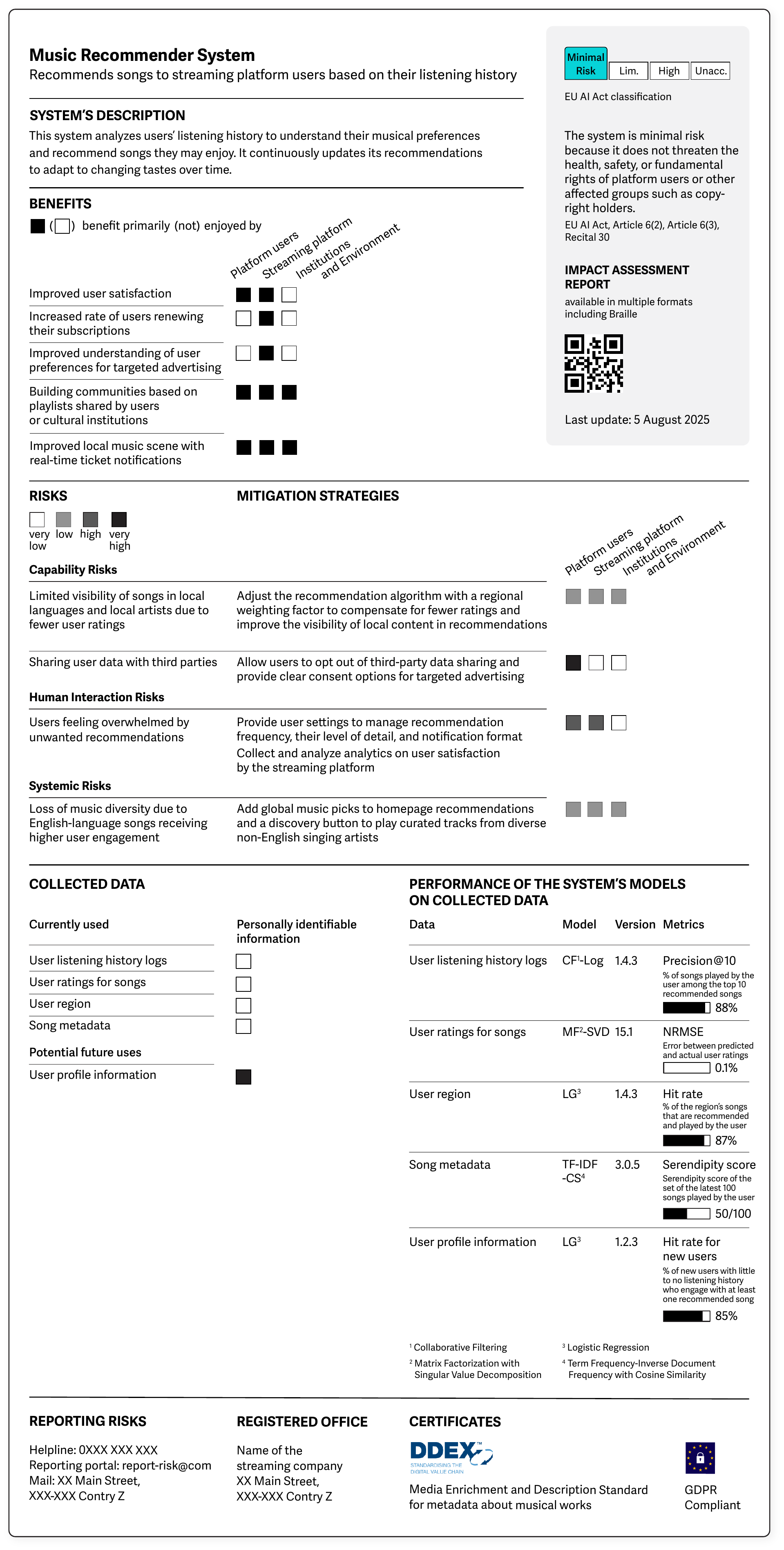}
   \caption{Final impact assessment card for a music recommender system that suggests songs to platform users based on their listening history. Also available at: \url{https://social-dynamics.net/ai-risks/impact-card/}.}
  \label{fig:card-music-recommender}
\end{figure*}

\begin{figure*}[h!]
  \centering
  \includegraphics[width=0.62\textwidth]{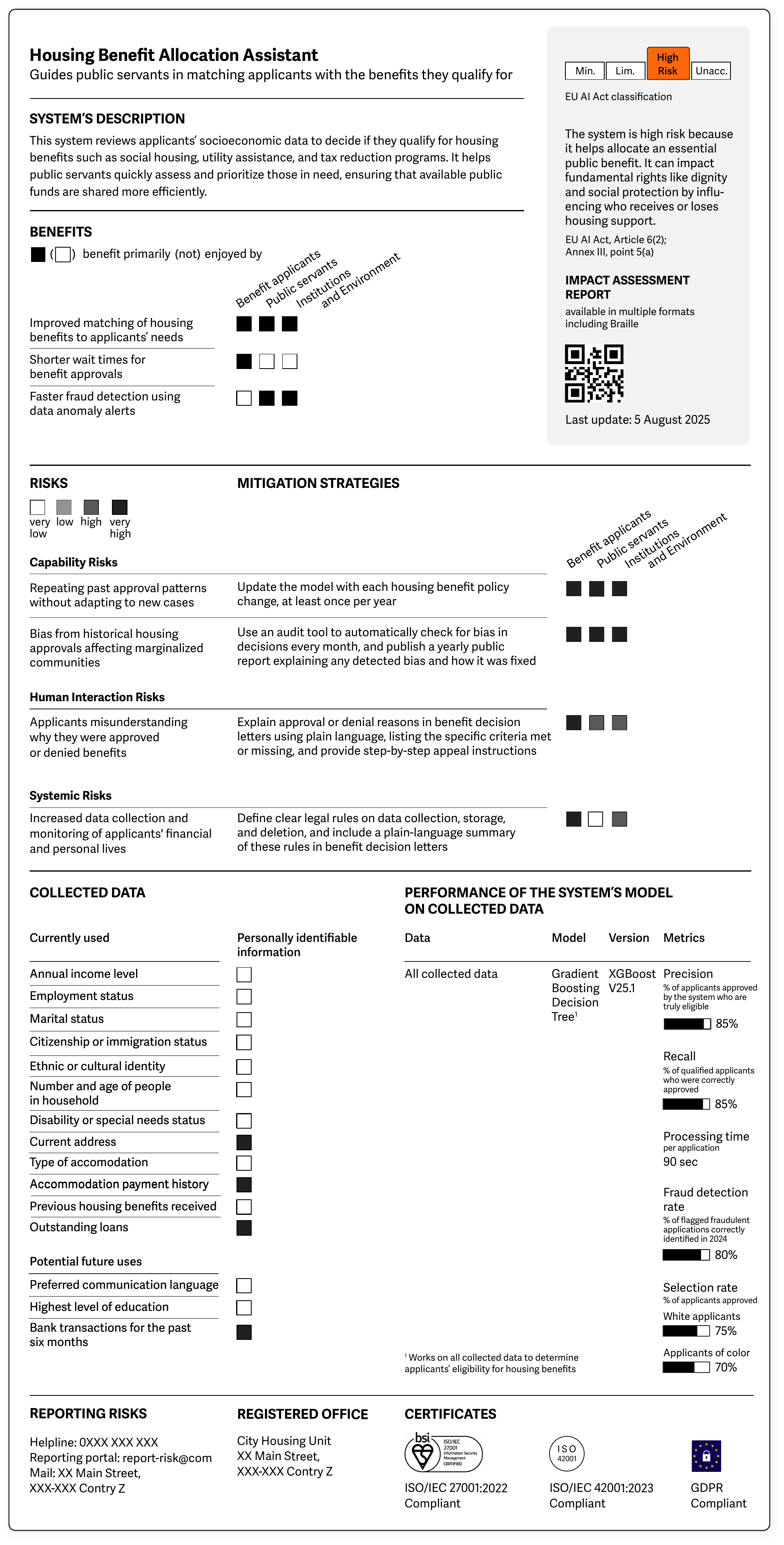}
  \caption{The impact assessment card for a housing benefit allocation assistant for public servants reviews applicants' socioeconomic data to match them with eligible benefits such as utility assistance and tax reduction programs. Also available at: \url{https://social-dynamics.net/ai-risks/impact-card/}.}
  \label{fig:card-benefit-assistant}
\end{figure*}

\end{document}